\documentclass[manuscript,preprint]{acmart}

\AtBeginDocument{%
  }

\usepackage{multirow}
\usepackage{graphicx}
\usepackage[normalem]{ulem}
\useunder{\uline}{\ul}{}

\begin{document}

\title{Predicting Selection Intention in Real-Time with Bayesian-based ML Model in Unimodal Gaze Interaction}

\author{Taewoo Jo}
\email{twj5349@yonsei.ac.kr}
\affiliation{%
  \institution{Yonsei University}
  \city{Seoul}
  \country{Republic of Korea}
}

\author{Ho Jung Lee}
\email{dearshwan@yonsei.ac.kr}
\affiliation{%
  \institution{Yonsei University}
  \city{Seoul}
  \country{Republic of Korea}
}

\author{Sulim Chun}
\email{slchun@yonsei.ac.kr}
\affiliation{%
  \institution{Yonsei University}
  \city{Seoul}
  \country{Republic of Korea}
}

\author{In-Kwon Lee}
\authornote{Corresponding Author}
\email{iklee@yonsei.ac.kr}
\affiliation{%
  \institution{Yonsei University}
  \city{Seoul}
  \country{Republic of Korea}
}

\renewcommand{\shortauthors}{Jo et al.}

\begin{abstract}
Eye gaze is considered a promising interaction modality in extende reality (XR) environments. However, determining selection intention from gaze data often requires additional manual selection techniques. We present a Bayesian-based machine learning (ML) model to predict user selection intention in real-time using only gaze data. Our model uses a Bayesian approach to transform gaze data into selection probabilities, which are then fed into an ML model to discriminate selection intentions. In Study 1, our model achieved real-time inference with an accuracy of 0.97 and an F1 score of 0.96. In Study 2, we found that the selection intention inferred by our model enables more comfortable and accurate interactions compared to traditional techniques.

\end{abstract}

\ccsdesc[300]{Human-centered computing~Interaction techniques}

\keywords{extended reality (XR), eye gaze, interaction, user interface (UI), intent prediction}

\begin{teaserfigure}
 \includegraphics[width=\textwidth]{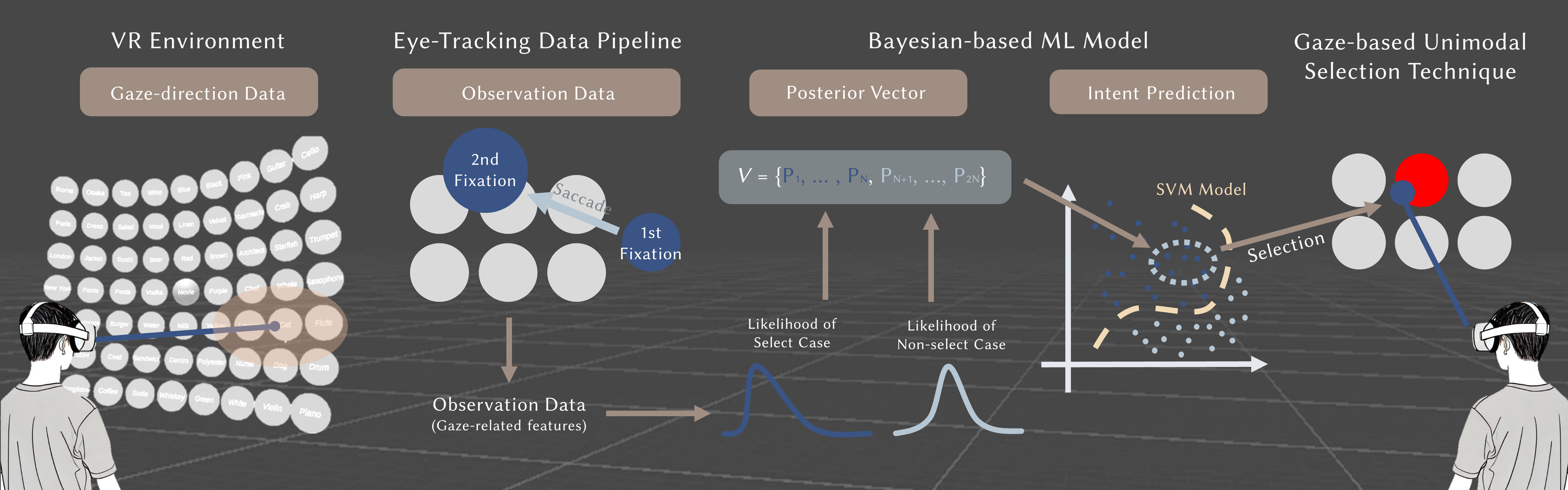}
     \caption{We propose a Bayesian-based machine learning (ML) model that predicts selection intention and a novel interaction technique that uses only gaze-based data, eliminating the need for additional manual selection in 3D target acquisition. First, our model processes the user's gaze data captured while observing a 3D target through an eye-tracking data pipeline to convert it into gaze-related features (observation data). We then use a Bayesian model to transform the observation data into a posterior vector. The posterior vector is then fed into an ML model, which infers whether the user intends to select the target. The our interaction technique uses the inference from our model to automatically select the target without any additional manual selection.}
 \label{fig:teaser}
\end{teaserfigure}


\maketitle

\section{Introduction}

Selecting a target is one of the fundamental tasks in human-computer interaction (HCI) enabling users to perform various interactions. Since fast and accurate pointing and selection techniques significantly enhance the interaction experience, the development of novel interaction techniques continues to be an active research area in HCI \cite{grossman2005:pt:2d, bi2013:touch:btc, moon2024:cont:btc:intpre:taskdesign:slc, sidenmark2023:gaze:head:ctrl}. Typically, handheld controllers have been used as the primary modality in VR/AR environments due to their ability to provide stable target selection performance. In addition, other modalities such as hand gesture tracking \cite{shi2023:gaze:hand:slc, zhu2023:hand:slc:pinch}, head movements \cite{choi2022:gaze:head:slc, sidenmark2023:gaze:head:ctrl} are also used for pointing and selection. However, these interaction techniques can cause physical fatigue, such as the "gorilla arm" effect in case of mid-air interaction \cite{hincapie2014:hand:ctrl:effort}, and limit natural hand movement.

With the advent of head-mounted displays (HMDs) capable of eye-tracking, gaze-based pointing and selection has attracted considerable interest. As an interaction modality, gaze offers the advantage of being hands-free, requiring minimal physical effort, and allowing for rapid pointing by allowing simultaneous observation and pointing \cite{sidenmark2023:gaze:head:ctrl, vertegaal2008:gaze, tanriverdi2000:gaze:hand}. However, using gaze as a standalone input modality for target acquisition presents several challenges. A primary issue is the difficulty of expressing selection confirmation through gaze alone, which limits its effectiveness as a unimodal input \cite{jacob1990:gaze:midas, choi2022:gaze:head:slc}. To address these issues, two approaches have been proposed in previous studies: (1) multimodal selection \cite{wei2023:gaze:pt:btc, choi2022:gaze:head:slc, shi2023:gaze:hand:slc}, in which gaze is used only for pointing and a controller key or specific hand gesture is used for selection, and (2) unimodal selection with additional procedures, such as dwell time, in which the user confirms a selection by fixating on the target for a predefined period of time longer than normal pointing \cite{narkar2024:gaze:dwell:slc:feat:intpre, isomoto2022:gaze:dwell:slc:intpre, yi2022:gaze:slc:eyelid}. However, when utilizing a hand-held controller or hand gestures, long-term hand usage can lead to physical fatigue \cite{boring2009:effort, hincapie2014:hand:ctrl:effort}. Similarly, the dwell method, which requires unnatural eye movements, can also cause similar issues \cite{duchowski2018:dwell:effort}.

To overcome this limitation, predicting the user selection intention has been explored, eliminating the need for manual selection step. While attempts have been made to apply this in 2D screen environments \cite{li2021:gaze:2d:btc, isomoto2018:gaze:dwell:slc:intpre, isomoto2022:gaze:dwell:slc:intpre}, the approach has been reported to suffer from usability issues due to the Midas touch problem and low pointing accuracy with gaze \cite{jacob1990:gaze:midas}. In VR environments, there have also been studies using gaze-based features to predict selection intention \cite{narkar2024:gaze:dwell:slc:feat:intpre, david2021:gaze:feat:slc}. However, more research is needed to effectively apply these methods in real-world VR environments.

In this paper, we present an Bayesian-based ML model that predicts the user selection intention by using selection probabilities derived from the Bayesian model (see Fig.~\ref{fig:teaser}). Our method uses a Bayesian model to convert gaze-related features into posterior vectors, which are rationale for predicting selection intention. As a result, our model can more accurately predict the user's selection intention. Furthermore, the model allows for real-time inference and enables gaze-based unimodal interaction without manual selection, which is more accurate  convenient than two-step selection techniques that additionally use controllers or dwell methods.

We evaluated our selection intention prediction model from two main perspectives: (1) its ability to accurately discriminate the user selection intention in real time, and (2) its effectiveness in supporting comfort and accurate target selection when implemented in a VR environment. In Study 1, we designed an experiment in which 20 participants performed a 3D target acquisition task to collect gaze data. We then built a Bayesian model based on the distribution of gaze-related features, and used the inferred selection probability data to train and evaluate a ML model. The prediction performance was found to be highly effective, with an accuracy of 0.97, an F1 score of 0.96, and an inference time of less than 1 ms. Thus, our model can quickly and accurately discriminate the user's selection intent. In Study 2, we compared the gaze-based interaction technique using our model with manual selection techniques such as controller and dwell methods in a 3D target selection environment. Our interaction technique allowed for accurate selection compared to the controller and dwell methods, and it also demonstrated lower workload and physical demand scores. These results suggest that our model enables comfort and accurate interaction in 3D environments without the need for manual selection processes.

In summary, the contributions of this study are as follows:

\textbf{Method for predicting selection intention in real time using only gaze data}: This study presents the most lightweight and accurate model for predicting user selection intention using a unimodal dataset based solely on gaze data.

\textbf{Improving the user experience of selection interactions using gaze-based unimodal interaction techniques}: We have demonstrated that our model, which differentiates selection intentions using only gaze data, enables the completion of 3D target selection tasks without the need for manual selection techniques. Moreover, this approach improves both performance and usability compared to traditional interaction techniques.

\section{Related Works}

\subsection{Pointing Technique}

Natural and efficient pointing and selection techniques are essential for providing optimal XR user experiences. Among these, the pointing task, which precedes selection, requires a technique that enables precise, fast targeting with minimal physical burden \cite{grossman2005:pt:2d, sidenmark2023:gaze:head:ctrl}. The human-computer interaction community has developed methods to predict pointing performance based on Fitts' Law \cite{fitts1954:fitts, mackenzie1992:fitts} and has proposed various pointing techniques applicable in 2D screen environments and 3D immersive environments \cite{mackenzie1992:fitts, grossman2005:pt:2d, wei2023:gaze:pt:btc}. Among these, pointing techniques using controllers as a modality have become the standard in 3D environment for interaction due to their high accuracy and satisfactory performance \cite{moon2024:cont:btc:intpre:taskdesign:slc, lu2020:cont:slc:taskdesign, yu2019:ctrl:pt:btc}. Furthermore, pointing techniques utilizing hand tracking and motion capture technologies have been proposed, leveraging hand gestures (or mid-air interactions) \cite{shi2023:gaze:hand:slc, zhu2023:hand:slc:pinch, periverzov2015:hand:slc}. However, challenges related to the accuracy of sensing and persistent issues related to the physical effort remain \cite{sidenmark2023:gaze:head:ctrl, hincapie2014:hand:ctrl:effort}. Recently, gaze-based pointing techniques have gained attention due to their ability to facilitate rapid and natural pointing movements with minimal physical effort \cite{sidenmark2023:gaze:head:ctrl, choi2022:gaze:head:slc, ren2024:gaze:hand:btc}. Despite the inherent noise in gaze data that can reduce accuracy, improvements have been made using the Bayesian criterion to enhance pointing precision \cite{li2021:gaze:2d:btc, wei2023:gaze:pt:btc, ren2024:gaze:hand:btc}.

\subsection{Selection Technique}

A convenient and accurate selection technique is as important as the pointing technique for seamless target interaction. Controllers enable easy selection confirmation through button presses, which, combined with satisfactory pointing performance, have become the standard in interaction techniques. Similarly, interaction techniques that utilize motion tracking of the hands are noteworthy for accurately performing selection tasks and offer the advantage of being hands-free compared to controller-based techniques \cite{shi2023:gaze:hand:slc, zhu2023:hand:slc:pinch}. However, both hand and controller use still fail to fully resolve the issue of physical effort during the pointing process.

Although gaze offers superior pointing performance, it has not been widely adopted as a uni-modal interaction technique, unlike controllers or hands. This is largely due to the Midas Touch problem, where unintended selections occur when gaze is used as a selection modality \cite{jacob1990:gaze:midas}. To prevent this, multimodal interaction techniques that express selection intention using other modalities (e.g., a button on a controller or a click gesture of the hand) have been proposed \cite{shi2023:gaze:hand:slc, choi2022:gaze:head:slc}. These methods are faster and more accurate than unimodal interaction techniques that use only a controller or hand \cite{sidenmark2023:gaze:head:ctrl}, but they come with the trade-off of being complex and unnatural to use, and they require a considerable amount of time to become accustomed to \cite{wei2023:gaze:pt:btc}.

Several techniques have been proposed that enable expressing selection intention using gaze alone. Notable among these are dwell time detection (DTD) \cite{jacob1990:gaze:dwell:slc, isomoto2018:gaze:dwell:slc:intpre}, eye blink \cite{yi2022:gaze:slc:eyelid, lu2021:blink}, and smooth pursuit \cite{vidal2013:pursuits}. These methods strive to provide a natural and efficient way to utilize gaze for selection without the need for additional physical inputs, aiming to overcome the limitations associated with the Midas Touch problem by ensuring that selections are intentional and precise. DTD is a selection technique proposed to avoid the Midas Touch problem, which involves unintentional selections when gaze is used as a selection modality. In this method, a user fixes their gaze on a target for a threshold time, often set in milliseconds (about 600 ~ 1200 ms), to indicate a selection intention \cite{jacob1990:gaze:dwell:slc, isomoto2018:gaze:dwell:slc:intpre, isomoto2022:gaze:dwell:slc:intpre}. The dwell threshold varies depending on the type of task and may also be adaptively adjusted based on the user's gaze movement patterns \cite{narkar2024:gaze:dwell:slc:feat:intpre}. Despite these efforts, the DTD method has not fully resolved the Midas Touch problem and still exhibits lower accuracy compared to other selection techniques \cite{choi2022:gaze:head:slc}. Additionally, the requirement to maintain gaze for a set duration to express intention can lead to reduced task performance, visual fatigue, and an unnatural user experience \cite{duchowski2018:dwell:effort}. These techniques often struggle to balance between being sensitive enough to allow for quick and easy selections and robust enough to prevent unintended interactions, thereby complicating their practical implementation and user acceptance in immersive environments.

\subsection{Selection Intention Prediction with Gaze Dynamics}

Gaze data is recognized as a promising source of information for predicting user interaction intentions. By utilizing gaze data, not only can low-level analysis reveal where users are looking, but it also provides insights into human attention and cognitive load concerning those areas \cite{krejtz2016:cok, duchowski2019:gaze:feat:cok, bednarik2012:gaze:intpre, li2021:gaze:2d:btc, david2021:gaze:feat:slc, narkar2024:gaze:dwell:slc:feat:intpre}. Research by David et al. \cite{david2021:gaze:feat:slc} has confirmed that various gaze-related features derived from analyzing gaze data can predict selection intentions within XR environments. Furthermore, Narkar et al. \cite{narkar2024:gaze:dwell:slc:feat:intpre} have demonstrated that using time-series gaze data and features as inputs in a long short-term memory model enables the real-time prediction of user selection intentions with high accuracy.

These findings suggest the feasibility of creating natural selection techniques in XR environments based on predicted intentions using gaze data. This approach leverages the deep, inherent connections between where users look and their underlying cognitive processes, thereby facilitating a more intuitive and less intrusive interaction techniques. Such advancements underscore the potential of gaze-based unimodal interactions to evolve beyond traditional techniques, offering a seamless integration of user intent recognition into immersive experiences.

\section{Bayesian Model for Selection Intention Prediction Problem}

We aim to address the gaze-based selection intention prediction problem by applying Bayes' theorem, using gaze-related features (observation data) to compute the selection intention probability (posterior data) and thereby infer the user intention. Our method can accurately predict selection intention in real-time, making it highly effective for environments where target selection tasks are performed using only gaze, without the need for a controller or hand gestures.

\subsection{Intent Prediction with Single Observation}

The process of predicting user intent based on Bayes' theorem using a single observation data point is as follows:

Let $S$ represent the state of the user, with $N$ possible states. These states could include intended target in 2D \cite{bi2013:touch:btc, zhai2012:btc, li2021:gaze:2d:btc} or 3D \cite{moon2024:cont:btc:intpre:taskdesign:slc, wei2023:gaze:pt:btc, yu2019:ctrl:pt:btc, ren2024:gaze:hand:btc} target selection scenario. Given observation data $O$, the probability that the user has selected a particular state $S$, denoted as $P(S\mid O)$, 
is calculated as follows:
\begin{equation}
\label{eq1}
    P(S\mid O) = \frac{P(O\mid S) \cdot P(S)}{P(O)}
\end{equation}
 
To infer the user-intended state, $S^*$, we identify the state with the highest probability:

\begin{equation}
\label{eq2}
    S^* = \arg\max_{S} P(S\mid O)
\end{equation}

According to Bayes' rule, $S^*$ can be expressed as:

\begin{equation}
\label{eq3}
    S^* = \arg\max_{S} \frac{P(O\mid S) \cdot P(S)}{P(O)}
\end{equation}
 
Here, $P(O\mid S)$ is the likelihood of observing $O$ given that the user is in state $S$, and $P(O)$ and $P(S)$ represent the probabilities of observing $O$ and selecting state $S$, respectively, both typically treated as constants. Thus, the intended state $S^*$ can be determined as:

\begin{equation}
\label{eq4}
    S^* = \arg\max_{S} {P(O\mid S)}
\end{equation}

\subsection{Selection Intention Prediction with Single Observation}

In the selection intention prediction problem, the state can be divided into two categories: Select (or True, $T$) and Non-Select (or False, $F$). Therefore, the user selection intention can be expressed as follows:

\begin{equation}
\label{eq5}
    S^* = 
    \begin{cases}
    T, & \text{if } P(O \mid T) > P(O \mid F) \\
    F, & \text{if } P(O \mid T) < P(O \mid F)
    \end{cases}
\end{equation}
 
This can be understood as a process of classification by calculating the posterior probabilities for both the true (Select) and false (Non-Select) cases. Thus, to predict the user selection intention, it is necessary to know the probability distributions for both cases — when the user selects and when the user does not select — based on a single observation.

\subsection{Selection Intention Prediction with Multiple Observations}

When the observation data consists of multiple observations ($M$ observations), then the comparison in Equation \ref{eq5} must be performed across $M$ dimensions. Assuming that all $M$ observations are independent, Bayes' theorem can be extended as follows:

\begin{equation}
P(T \mid O_1, O_2, \dots, O_M) = \frac{P(O_1 \mid T) \cdot P(O_2 \mid T) \cdots P(O_M \mid T) \cdot P(T)}{P(O_1) \cdot P(O_2) \cdots P(O_M)}
\end{equation}
 
This approach is analogous to the Naive Bayes method used in classification problems. However, the gaze-related features that we intend to use as observation data in this study are not independent of each other \cite{krejtz2016:cok, david2021:gaze:feat:slc}. Although some studies have used Bayesian models in the form of relative probabilities to address this issue \cite{wei2023:gaze:pt:btc}, this method is difficult to generalize to more than three observations. Therefore, in this study, we approach the problem as an $M$-dimensional classification problem. 

For $M$ observation data points, let $F_{T}(O)$ and $F_{F}(O)$ be the likelihood functions for the true and false cases, respectively. We define the posterior vector $\textbf{V}$ as follows:

\begin{equation}
\mathbf{V} = \left( F_{T}(O_1), F_{T}(O_2) \dots, F_{T}(O_M), F_{F}(O_1), F_{F}(O_2) \dots, F_{F}(O_M) \right)
\end{equation}

Suppose that we know the actual class (\textit{True} or \textit{False}) of the computed 2$M$-dimensional posterior vector. This format can be utilized as input for the Support Vector Machine (SVM) method, which is optimized for classifying vector data. SVM is known for its fast inference time, and when trained with a Gaussian kernel, it can accurately and efficiently discriminate high-dimensional vectors.

Thus, the primary goal of our first study is to develop a Bayesian model that converts multiple observations into a posterior vector, and an SVM model that classifies the user selection intention based on this posterior vector.

\section{Study 1: Construction and Evaluation of the Selection Intention Prediction Model}
\label{sec:study1}

The aim of Study 1 was to develop an Bayesian-based ML model capable of predicting selection intentions from gaze data in real time. The study consisted of three main steps: (1) a user study was conducted in which participants performed target acquisition tasks to collect gaze data; (2) the observed data were fitted to a probability density function (pdf) to construct the Bayesian model; and (3) using the posterior vector derived from the Bayesian model, an ML model was trained and evaluated for its performance in predicting selection intentions based on gaze data.

\subsection{Data Collection}
\label{subsec:data_collection_study1}

This subsection outlines the implementation and procedure of the user study conducted to collect gaze data. The study was structured to collect gaze data in different scenarios by differentiating between the configuration of the targets and the complexity of target acquisition. This method ensured comprehensive coverage of different conditions.

\subsubsection{Participants and Apparatus}
\label{subsubsec:participants_apparatus_study1}

We recruited 20 participants from the local area near our research site for data collection, targeting individuals with normal or corrected-to-normal vision. Only participants using non-glasses corrective devices were eligible, as glasses could interfere with HMD wear or eye-tracking functionality. In addition, individuals with color vision deficiencies, such as color blindness, were excluded from the study. Twelve participants reported previous experience with VR HMDs.
The VR environment was developed using Unity 2022.3.16f1 and experienced through the Meta Quest Pro HMD. Eye tracking data was recorded at a frequency of 60-66 Hz.

\subsubsection{Task design}
\label{subsubsec:task_design_study1}

\begin{figure}[ht]
    \centering
    \includegraphics[width=0.5\linewidth]{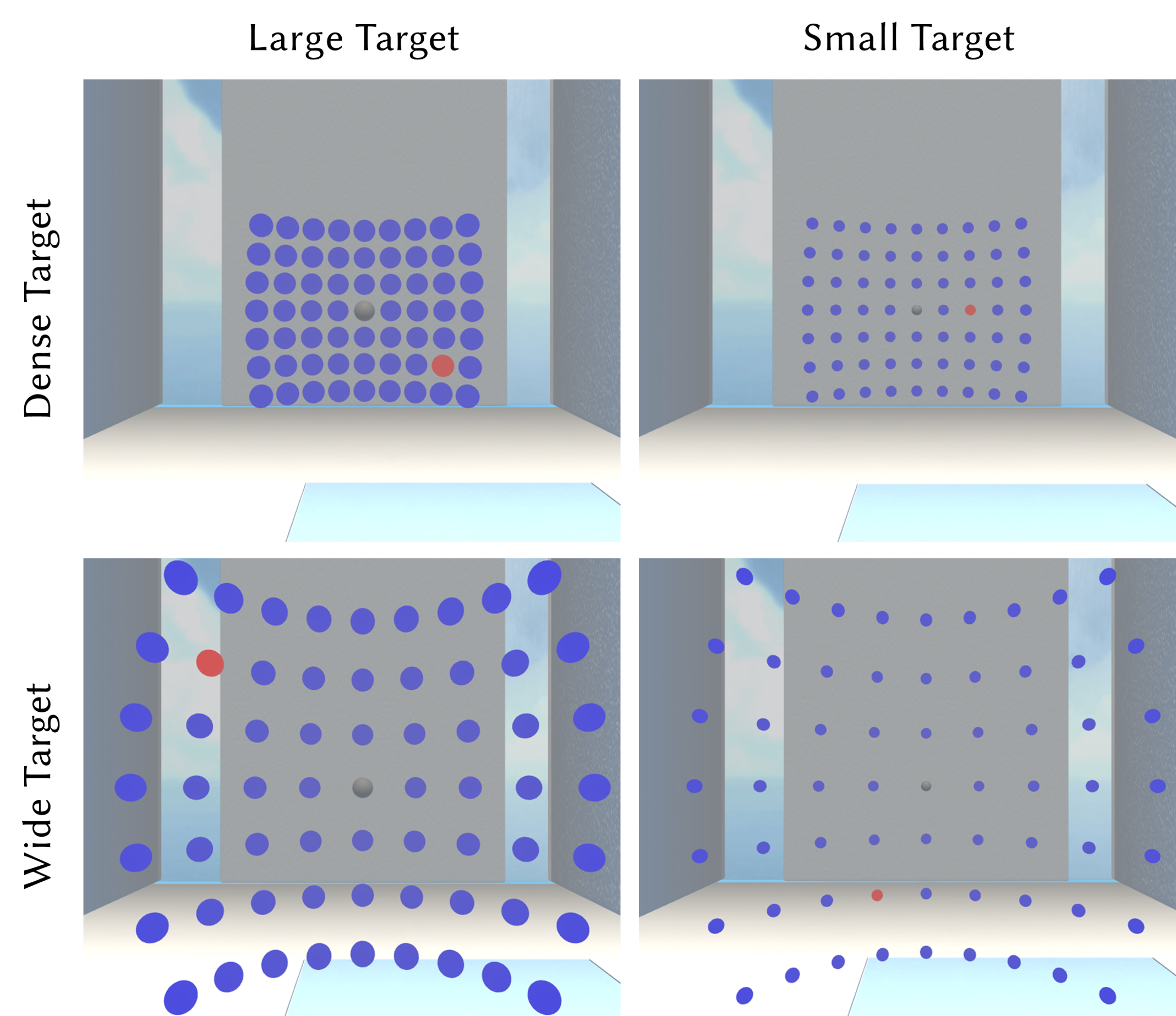}
    \caption{We configured a total of four types of target configurations by combining two factors: target size (\textit{Large} or \textit{Small}) and target density (\textit{Wide} or \textit{Dense}).}
    \label{fig:config}
\end{figure}

We implemented a modified version of the target acquisition task used in previous research \cite{lu2020:cont:slc:taskdesign, moon2024:cont:btc:intpre:taskdesign:slc}. In this task, spheres are arranged in a $9 \times 7$ grid, with a central start object (colored gray) surrounded by distractors (colored blue). Among the distractors, objects located between the outermost layer and adjacent to the starting object serve as potential target objects (26 in total) and are randomly activated as task targets. Additionally, four configurations were created by varying two levels of target density (\textit{Dense} and \textit{Wide}) and two target sizes (\textit{Small} and \textit{Large}) (see Fig.~\ref{fig:config}). In the \textit{Dense} target condition, objects are spaced 7.5 degrees apart, whereas in the \textit{Wide} condition, the spacing is 15 degrees. The \textit{Small} target width is 3 degrees, while the \textit{Large} target width is 6 degrees.

\begin{figure}[ht]
    \centering
    \includegraphics[width=0.5\linewidth]{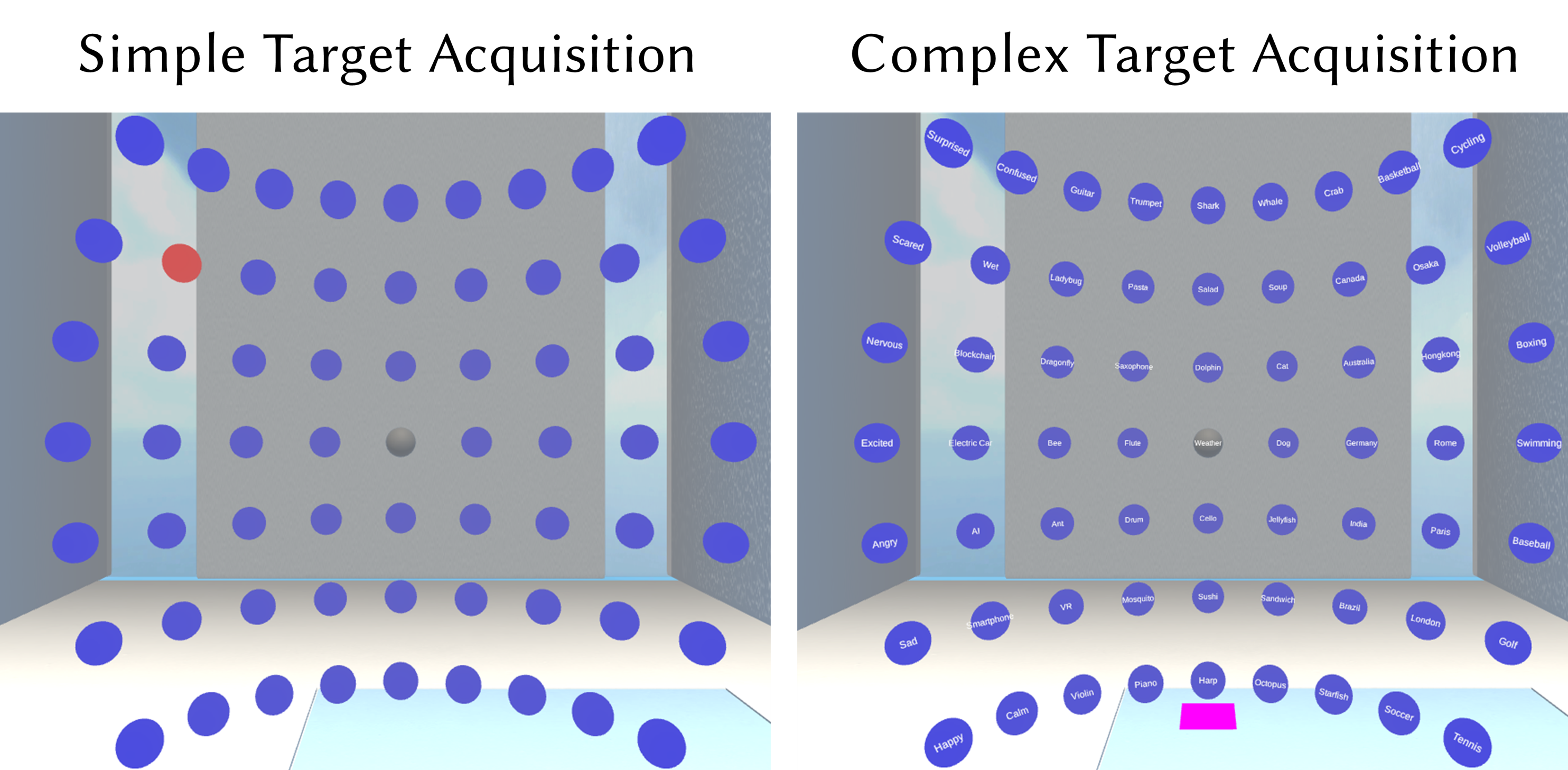}
    \caption{We categorized acquisition complexity into two levels based on the difficulty of identifying the target. In the \textit{Simple} acquisition task (left), a red target object is placed among blue distractors, and the task is to click on the target. In the \textit{Complex} acquisition task (right), each distractor has a random word written on it, and the task is to find and click the target object that has a word belonging to the same category as the word written on the starting object.}
    \label{fig:acqu}
\end{figure}

In order to present a variety of task difficulties, two different task types were designed. The first, a simple acquisition task (or \textit{Simple}), is similar to the task used in previous studies \cite{lu2020:cont:slc:taskdesign, moon2024:cont:btc:intpre:taskdesign:slc}. When the user points to the starting object, one of the potential target objects is randomly activated to become the target object and turns red. After accurately pointing and confirming the selection of the target object, it returns to blue, accompanied by auditory feedback. The task is reset by pointing at the start object again, which activates a new target object, and the task is completed after interacting with all potential target objects.

The second task is a complex acquisition task (or \textit{Complex}) in which a category word is displayed on the starting object, and the potential target objects (still colored blue, not red) contain words belonging to that category, while the distractor objects show unrelated words (see Fig.~\ref{fig:acqu}). At the start of the task, participants were asked to locate and select the object containing the correct category word. After accurately pointing and confirming the selection, the target object turns red and a sound feedback is played. After pointing again at the starting object, a new category word appears and the words on all objects are shuffled, with one potential target randomly selected as the new target. The task ends when all potential targets have been interacted with. The words used in the \textit{Complex} task were chosen from the language commonly used by the participants.

\subsubsection{Study design and procedure}
\label{subsubsec:study_design_procedure_study1}

The data collection study followed a $2 \times 2 \times 2$ within-subjects factorial design: \textit{Target Density} (\textit{Dense} and \textit{Wide}) $\times$ \textit{Target Size} (\textit{Small} and \textit{Large}) $\times$ \textit{Task Complexity} (\textit{Simple} and \textit{Complex}). Participants performed the task in eight different conditions, completing two tasks per condition, with each task consisting of 26 trials. In all conditions, gaze pointing was used as the pointing technique, and selection was confirmed by clicking the trigger button on the controller.

Prior to the start of the experiment, participants completed a demographic questionnaire. They then received instructions on the tasks, the VR HMD, and the use of the controller, followed by an explanation of the eye-tracking calibration protocol. After calibration, participants underwent a practice session in which they completed five trials for each of the eight conditions to familiarize themselves with the task. This was followed by the main session, in which the order of the conditions was counterbalanced using a counterbalanced Latin square \cite{bradley1958:latin}. After each task, participants completed a one-question discomfort survey on a 10-point Likert scale to assess their level of fatigue \cite{fernandes2016:survey}. If a score of 7 or higher was reported, a rest period was provided, and if a score of 10 was reported, the experiment was terminated. In addition, the experiment was terminated if the total duration exceeded 90 minutes to prevent degradation of gaze data quality due to accumulated fatigue. Participants who completed the entire study finished 416 trials. The study was conducted under institutional IRB approval.

\subsection{Gaze Data Pipeline}
\label{subsec:gaze_data_pipeline_study1}




This subsection explains how the gaze data collected in Section \ref{subsec:data_collection_study1} were processed and recorded. Our data pipeline converts eye and head tracking data into gaze direction data and computes gaze-related features when a fixation event is detected.

First, eye (i.e., gaze-in-head frame) and head (i.e., head-in-world frame) direction data are collected using the HMD's tracking function, and gaze direction data (i.e., gaze-in-world frame) are computed by quaternion multiplication. Then, both head and gaze direction data are converted to angular units to compute angular dispersion and velocity for each frame. These computed dispersion and velocity values are used in the I-VDT (Velocity and Dispersion Threshold Identification) algorithm to detect fixation and saccade events \cite{sendhilnathan2022:ivdt:error, david2021:gaze:feat:slc}. A gaze velocity greater than 70 degrees/second is classified as a saccade, while events with a velocity less than 30 degrees/second, a dispersion less than 1 degree, and a duration of at least 30 ms are classified as fixations.

When a fixation event is detected by the I-VDT algorithm, gaze-related features are computed within a data window as shown in Fig.~\ref{fig:teaser}. When a fixation is detected, features are computed for both the previous fixation (1st fixation) and the preceding saccade. Features are also computed for the currently detected fixation (2nd fixation). When the fixation ends, the calculation of gaze-related features is stopped. The list of computed features can be found in the Supplementary Material.

In addition, when a participant confirms a selection, this event is recorded along with the gaze data. Thus, for each frame, the recorded data include eye direction data, head direction data, gaze direction data, fixation event detection, gaze-related features, and selection confirmation status. All data transformations, calculations, and recording are performed in real time.

\subsection{Bayesian Model Construction}
\label{subsec:bayesian_model_construction_study1}



This subsection describes the process of constructing the Bayesian model that transforms the observation data into a posterior vector. In this study, only gaze-related feature data characterized by different distributions based on selection confirmation were included in the Bayesian model. For each condition, gaze-related features were classified according to selection confirmation, and features with different distributions between classes were selected as observation data. The distribution of these selected observation data was then fitted to a probability distribution function.

Classes were divided based on whether selection confirmation occurred, resulting in a select class (or \textit{True} class) and a non-select class (or \textit{False} class). Selection confirmation that occurred during a fixation state was classified into the \textit{True} class. In cases where a fixation ended without selection confirmation, the corresponding features at the end of the fixation were classified into the \textit{False} class.

Next, we investigated which gaze-related features exhibited distinct distributions across selection classes, and this process was performed for each condition. First, the data were categorized by condition type, feature type, and class type, and outliers were identified and treated using the interquartile range method. An independent $t$-test was then performed to determine if the different class types had different distributions. The class type (\textit{True} or \textit{False}) was used as a factor for each condition and feature. Features were selected as observation datas if significant differences (as indicated by a $p$-value for the class factor below 0.05) between the \textit{True} and \textit{False} classes were found in six or more of the eight conditions (see Table~\ref{tab:feature:selection}). The results of this analysis are provided in the Supplementary Materials.

\begin{table}[]
\caption{Each Feature Group (1st Fixation, 2nd Fixation, Saccade, Coefficient K) and their respective specific features. A count of statistically significant differences based on class type factor, confirmed via independent t-tests, was added. Features were selected when statistical significance was confirmed in 6 or more out of the 8 conditions. Selected features are marked with *.}
\label{tab:feature:selection}
\resizebox{\textwidth}{!}{%
\begin{tabular}{l|clllcllc}
\hline
\textbf{Feature Group} &
  \multicolumn{4}{c|}{\textbf{1st Fixation}} &
  \multicolumn{4}{c}{\textbf{2nd Fixation}} \\ \hline
\textbf{Feature} &
  \multicolumn{1}{l}{\begin{tabular}[c]{@{}l@{}}Fixation\\ duration\end{tabular}} &
  \begin{tabular}[c]{@{}l@{}}*Fixation\\ std x\end{tabular} &
  \begin{tabular}[c]{@{}l@{}}*Fixation\\ std y\end{tabular} &
  \multicolumn{1}{l|}{\begin{tabular}[c]{@{}l@{}}*Fixation\\ velocity\end{tabular}} &
  \multicolumn{1}{l}{\begin{tabular}[c]{@{}l@{}}*Fixation\\ duration\end{tabular}} &
  \begin{tabular}[c]{@{}l@{}}*Fixation\\ std x\end{tabular} &
  \begin{tabular}[c]{@{}l@{}}*Fixation\\ std y\end{tabular} &
  \multicolumn{1}{l}{\begin{tabular}[c]{@{}l@{}}*Fixation\\ velocity\end{tabular}} \\
\textbf{Count} &
  4 &
  \multicolumn{1}{c}{8} &
  \multicolumn{1}{c}{7} &
  \multicolumn{1}{c|}{8} &
  8 &
  \multicolumn{1}{c}{8} &
  \multicolumn{1}{c}{8} &
  8 \\ \hline
\textbf{Feature Group} &
  \multicolumn{7}{c|}{\textbf{Saccade}} &
  \textbf{Coefficient K} \\ \hline
\textbf{Feature} &
  \multicolumn{1}{l}{\begin{tabular}[c]{@{}l@{}}Saccade\\ duration\end{tabular}} &
  \begin{tabular}[c]{@{}l@{}}Saccade\\ amplitude\\ eye\end{tabular} &
  \begin{tabular}[c]{@{}l@{}}*Saccade\\ amplitude\\ head\end{tabular} &
  \begin{tabular}[c]{@{}l@{}}*Saccade\\ amplitude\\ gaze\end{tabular} &
  \multicolumn{1}{l}{\begin{tabular}[c]{@{}l@{}}*Saccade\\ velocity\\ eye\end{tabular}} &
  \begin{tabular}[c]{@{}l@{}}*Saccade\\ velocity\\ head\end{tabular} &
  \multicolumn{1}{l|}{\begin{tabular}[c]{@{}l@{}}Saccade\\ velocity\\ gaze\end{tabular}} &
  \multicolumn{1}{l}{*Coefficient K} \\
\textbf{Count} &
  5 &
  \multicolumn{1}{c}{5} &
  \multicolumn{1}{c}{8} &
  \multicolumn{1}{c}{6} &
  8 &
  \multicolumn{1}{c}{8} &
  \multicolumn{1}{c|}{5} &
  6 \\ \hline
\end{tabular}%
}
\end{table}

Finally, the selected observation data were fitted to different functions. Maximum likelihood estimation was used to fit the data to 15 different probability density functions, including the normal, Weibull, and Gamma distributions. The Kolmogorov-Smirnov test was then applied to select density functions with a $p$-value greater than 0.05. In cases where multiple density functions met this criterion, the one with the highest $p$-value was selected. If no suitable fitting function was found, a uniform distribution was applied to avoid affecting the ML model's classification. Fitting results are summarized in the Supplementary Material.

\subsection{Training and Evaluating ML Model}

This section describes the process of training the ML model using the posterior vector computed from the Bayesian model, as well as the classification performance. In addition, the efficiency of the proposed Bayesian-based ML model is demonstrated by comparing it with previous studies that used gaze-related features \cite{narkar2024:gaze:dwell:slc:feat:intpre, isomoto2022:gaze:dwell:slc:intpre, david2021:gaze:feat:slc} and with ML models trained directly on observation data without the Bayesian model.

The ML model was tasked with classifying whether the posterior vector belonged to the \textit{True} or \textit{False} class. A Gaussian kernel-based SVM model was employed. Due to the class imbalance in the posterior vector dataset (with a ratio of 1:4 for the \textit{Simple} condition and 1:80 for the \textit{Complex} condition), the dataset was sampled to achieve a class ratio of 1:3. The dataset was split into training, validation, and test sets in a ratio of 6:2:2.

\begin{table}[]
\caption{Model performance based on the inclusion of each Feature Group. The highest performance was observed when excluding only the features from the Saccade group.}
\label{tab:model:ablation}
\begin{tabular}{lllll}
\hline
                & Acc   & F1    & AUC-ROC & Inference Time (ms) \\ \hline
All Features    & 0.965 & 0.947 & 0.979   & 0.036               \\
No 1st Fixation & 0.823 & 0.716 & 0.834   & 0.047               \\
No 2nd Fixation & 0.965 & 0.947 & 0.981   & 0.032               \\
\textbf{No Saccade}      & \textbf{0.972} & \textbf{0.953} & \textbf{0.988}   & \textbf{0.031}               \\
Observation Data      & 0.838 & 0.455 & 0.685   & 0.082               \\ \hline
\end{tabular}
\end{table}

To optimize the performance of the SVM model, hyperparameter tuning and an ablation study were conducted. For the SVM, the gamma value of the Gaussian kernel and the complexity parameter were adjusted. The ablation study evaluated the model performance under the following conditions: 1) using all posterior vectors, 2) excluding 1st fixation-related features, 3) excluding 2nd fixation-related features, 4) excluding saccade-related features and 5) observation data. Through these steps, the selection intention prediction performance of the proposed Bayesian-based ML model was summarized in Table \ref{tab:model:ablation} for each condition.

\section{Study 2: Evaluation of the Gaze-based Unimodal Interaction Technique}

The second study evaluates the efficiency and effectiveness of the gaze-based unimodal interaction technique, utilizing the previously developed model. Specifically, we implemented a selection technique that determines whether to select the pointed target based on the inference results from the Bayesian-based ML model trained in Study 1. When combined with a gaze-based pointing technique, this approach enables target selection using only gaze modality, eliminating the need for manual selection confirmation.

\subsection{User Study}

The aim of Study 2 is to investigate the differences in interaction experience when performing target selection tasks using different selection techniques. Three selection techniques were implemented in the VR environment: our technique based on the model trained in Study 1 and two commonly used selection techniques. The following sections provide a detailed explanation of the user study design, the target acquisition tasks performed, the description of each interaction technique, and other relevant details.

\subsubsection{Participants}

We recruited 23 participants, none of whom had participated in Study 1, all with normal or corrected-to-normal vision. For participants with corrected-to-normal vision, only those using non-glasses corrective devices were eligible for the study. In addition, individuals with color blindness were excluded from participation. Of the participants, 15 reported previous experience with VR HMDs. The VR environment was implemented using the same methods as in Study 1, and the same VR HMD was used.

\subsubsection{Task and interaction techniques}

The task implemented in this user study was the \textit{Complex} target acquisition task used in Study 1. Participants had to read the category word displayed on the starting object and find the target object labeled with a word belonging to that category among the distractors. In contrast to Study 1, only two types of target configurations were used: \textit{Large} $\times$ \textit{Dense} and \textit{Small} $\times$ \textit{Wide}.

Three interaction techniques have been implemented, each using different pointing and selection methods as follows:
\begin{enumerate}
    \item \textit{Gaze}: Gaze-based pointing and selection based on the selection intention prediction model
    \item \textit{Dwell}: Gaze-based pointing and dwell-based selection
    \item \textit{Cont}: Controller-based pointing and selection
\end{enumerate}

Gaze-based pointing used the gaze direction data computed in Section \ref{subsec:gaze_data_pipeline_study1} using a raycast method to identify which object the user was pointing at. Controller-based pointing used the handheld controller of the VR HMD, also using a raycast method to determine which object the user was pointing at.

The selection mechanism for the \textit{Gaze} condition relied on gaze data processed through the pipeline, where the computed observation data was fed into the Bayesian-based ML model. Based on the model's inference, the system decided whether to select the pointed object. For the \textit{Dwell} condition, the pointed object was selected if the user maintained a fixation state for more than 600 ms. For the \textit{Cont} condition, the pointed object was selected when the trigger button on the handheld controller was pressed.

\subsubsection{Study design and procedure}

The user study followed a $2 \times 3$ within-subjects factorial design: \textit{Target Configuration} (\textit{Dense} $\times$ \textit{Large} and \textit{Wide} $\times$ \textit{Small}) by \textit{Interaction Technique} (\textit{Gaze}, \textit{Dwell}, and \textit{Cont}). Participants performed the Complex target acquisition task under six different conditions, completing one task per condition, with each task consisting of 26 selection trials.

Prior to the start of the experiment, participants completed a demographic questionnaire. They were then instructed on the task to be performed and how to use the VR HMD and controller. Instructions for the eye tracking calibration protocol were also provided. After completing the calibration, participants underwent a practice session in which they experienced five trials of each of the six conditions to familiarize themselves with the task. This was followed by the main session, in which the order of conditions was counterbalanced using a counterbalanced Latin square design \cite{bradley1958:latin}.

At the end of each task, participants completed a discomfort survey \cite{fernandes2016:survey}, the System Usability Scale (SUS) \cite{brooke1996:sus}, and the NASA-TLX \cite{hart1988:nasa} survey. If participants rated discomfort as 7 or higher, they were given a break, and the experiment was terminated if they rated discomfort as 10. The study was IRB approved.

\subsection{Results}

To examine the effects of the \textit{Target Configuration} and \textit{Interaction Technique} factors, we conducted a statistical analysis. First, the normality of the data was assessed using the Shapiro-Wilk test. If the data met the normality assumption, we proceeded with a two-way repeated measures analysis of variance (RM-ANOVA) to test for interaction effect and main effects. In addition, RM-ANOVA analyses were conducted separately for the \textit{Large $\times$ Dense} and \textit{Small $\times$ Wide} configurations to observe the effect of \textit{Interaction Technique} factor.

For all RM-ANOVA analyses, if Levene's test indicated a violation of the sphericity assumption, the degrees of freedom were adjusted accordingly. If the data violated the normality assumption, the Friedman test was used instead. When a significant interaction or main effect was observed by RM-ANOVA or the Friedman test, post-hoc pairwise comparisons were performed using the Bonferroni correction to adjust for multiple comparisons. In the following section, we highlight the results where statistical significance was found, while detailed statistical analysis results can be found in the supplementary material.

\subsubsection{Time-to Completion (TTC)}

TTC was measured as the time from the moment the participant pointed at the start object until the target object was selected. The mean TTC for each task was calculated by averaging the TTC across the 26 trials. A smaller TTC indicates faster target acquisition.

\begin{figure}[ht]
    \centering
    \includegraphics[width=\linewidth]{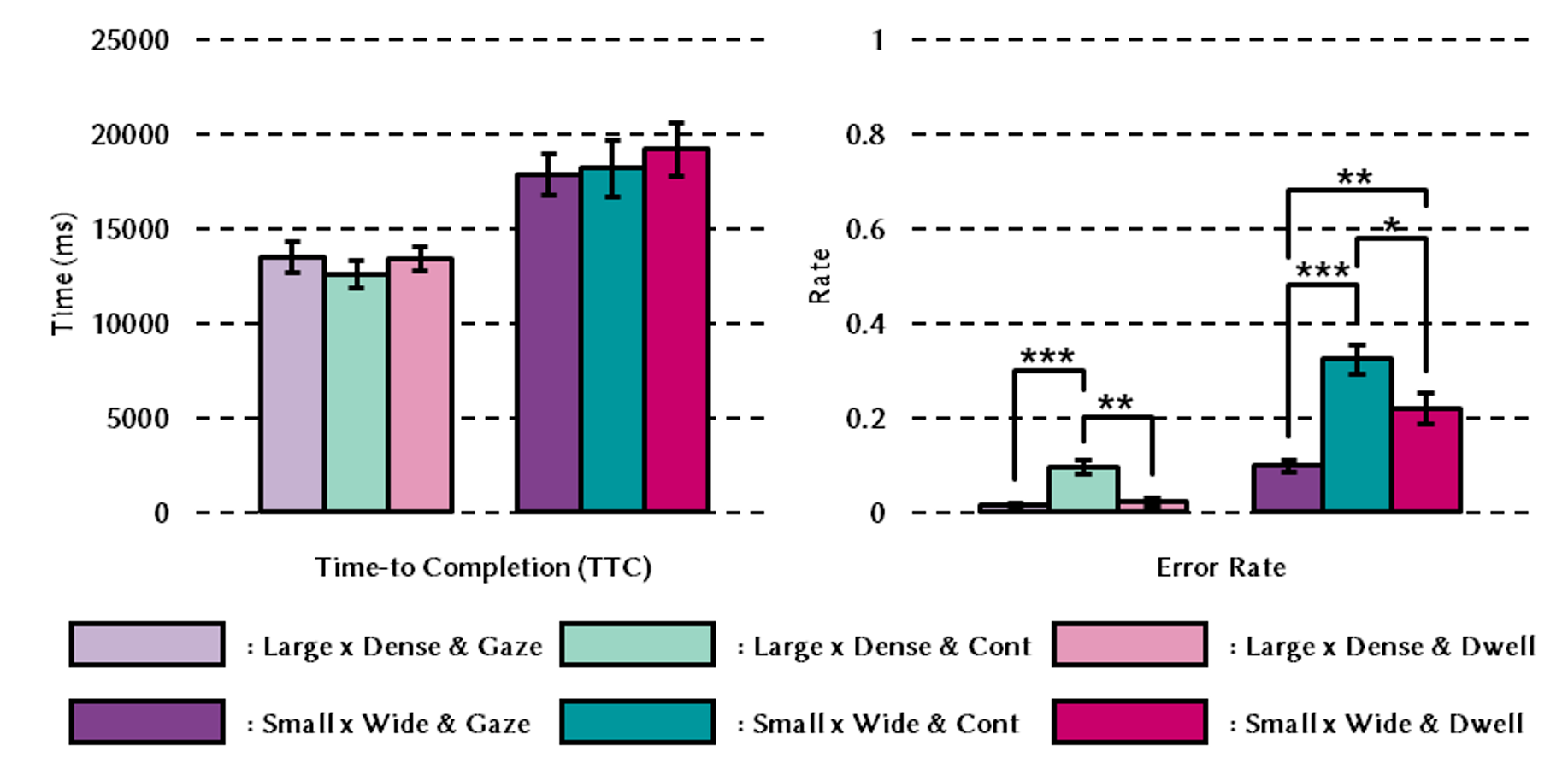}
    \caption{Time-to-Completion (TTC) (Left) and Error Rate (Right) based on target configuration and input modality. Statistically significant differences identified through post-hoc tests are indicated with connecting lines, where *, **, and *** represent p-values less than 0.05, 0.01, and 0.001, respectively.}
    \label{fig:performance}
\end{figure}

The two-way RM-ANOVA analysis of mean TTC revealed a significant main effect of configuration, while no significant interaction effect was observed (see Fig.~\ref{fig:performance} left). Furthermore, one-way ANOVA analyses for the \textit{Large $\times$ Dense} and \textit{Small $\times$ Wide} configurations showed no significant differences in TTC across selection techniques. This indicates that while the \textit{Target Configuration} had a significant effect on task completion time, the type of \textit{Interaction Technique} did not have a significant effect on task completion time.

\subsubsection{Error rate}

The error rate was calculated by dividing the number of unsuccessful clicks by the total number of clicks during the task. A higher error rate indicates that participants had more difficulty making selections due to the configuration or technique.

The two-way RM-ANOVA analysis of the error rate revealed a significant interaction effect as well as significant main effects for both configuration and technique (see Fig.~\ref{fig:performance} right). In addition, a one-way ANOVA for the \textit{Large $\times$ Dense} configuration revealed a significant difference in error rate based on the technique used. Post-hoc tests showed that the \textit{Cont} condition resulted in significantly higher error rates compared to both the \textit{Gaze} and \textit{Dwell} conditions.

For the \textit{Small $\times$ Wide} configuration, a significant difference in error rate between techniques was also observed. Post-hoc tests indicated that the \textit{Cont} condition had significantly higher error rates compared to both the \textit{Gaze} and \textit{Dwell} conditions, and the \textit{Gaze} condition had a significantly lower error rate than the \textit{Dwell} condition.

These results suggest that the differences in error rates based on configuration indicate that configuration influenced task difficulty. In addition, the \textit{Cont} condition produced more errors compared to \textit{Gaze} and \textit{Dwell} in both configurations, while the \textit{Gaze} condition produced the most fewest errors in the \textit{Small $\times$ Wide} configuration.

\subsubsection{NASA-TLX}

The NASA-TLX survey was used to measure task workload, with a higher overall score indicating greater workload. In addition, within the sub-dimensions of the survey, the physical demand score reflects how much physical strain the task imposed, with higher scores indicating more physical strain.

\begin{figure}[ht]
    \centering
    \includegraphics[width=\linewidth]{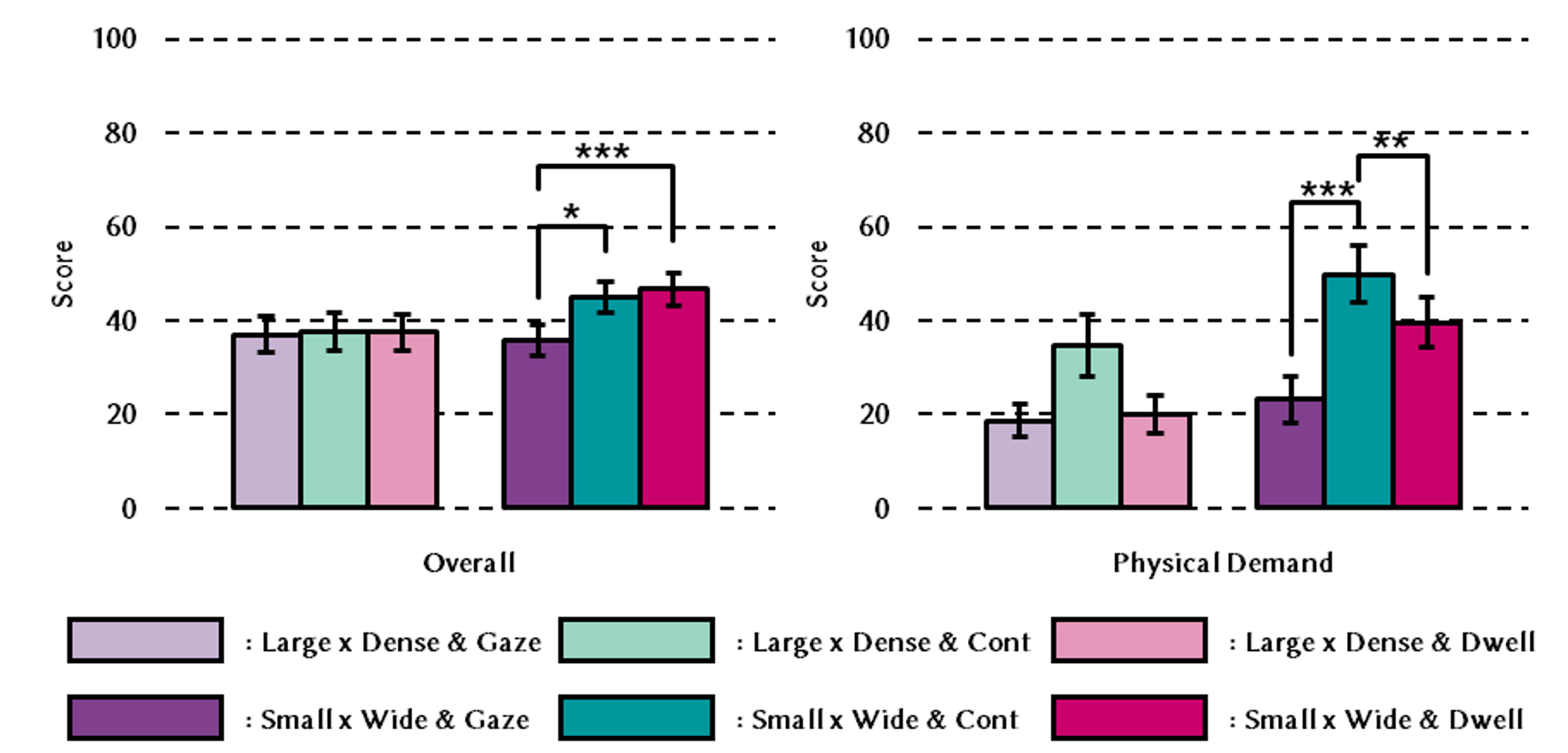}
    \caption{Overall NASA-TLX score (Left) and physical demand (Right) based on target configuration and input modality. Statistically significant differences identified through post-hoc tests are indicated with connecting lines, where *, **, and *** represent p-values less than 0.05, 0.01, and 0.001, respectively.}
    \label{fig:nasa}
\end{figure}

The two-way RM-ANOVA analysis of the overall score revealed both a significant interaction effect and a main effect of Interaction Technique (see Fig.~\ref{fig:nasa} left). In the \textit{Large $\times$ Dense} configuration, the one-way RM-ANOVA analysis for \textit{Interaction Technique} showed no statistically significant differences. However, in the \textit{Small $\times$ Wide} configuration, there was a significant difference in total score based on \textit{Interaction Technique}. Post-hoc tests revealed that the \textit{Gaze} condition had a significantly lower overall score compared to both the \textit{Cont} and \textit{Dwell} conditions. This suggests that while the workload was similar across techniques in the \textit{Large $\times$ Dense} configuration, the \textit{Gaze} condition resulted in less workload in the \textit{Small $\times$ Wide} configuration compared to the other techniques.

For the Physical Demand dimension, the two-way RM-ANOVA analysis revealed a significant interaction effect, as well as significant main effects for both configuration and technique (see Fig.~\ref{fig:nasa} right). One-way RM-ANOVA analyses for both the \textit{Large $\times$ Dense} and \textit{Small $\times$ Wide} configurations confirmed significant differences in physical demands scores across techniques. In particular, post-hoc tests showed that in the \textit{Small $\times$ Wide} configuration, the \textit{Gaze} condition had a significantly lower physical demand score compared to the \textit{Cont} and \textit{Dwell} conditions. This indicates that participants experienced greater physical demands with the \textit{Cont} techniques, especially in the \textit{Small $\times$ Wide} configuration.

\subsubsection{System usability scale (SUS)}

\begin{figure}[h]
    \centering
    \includegraphics[width=\linewidth]{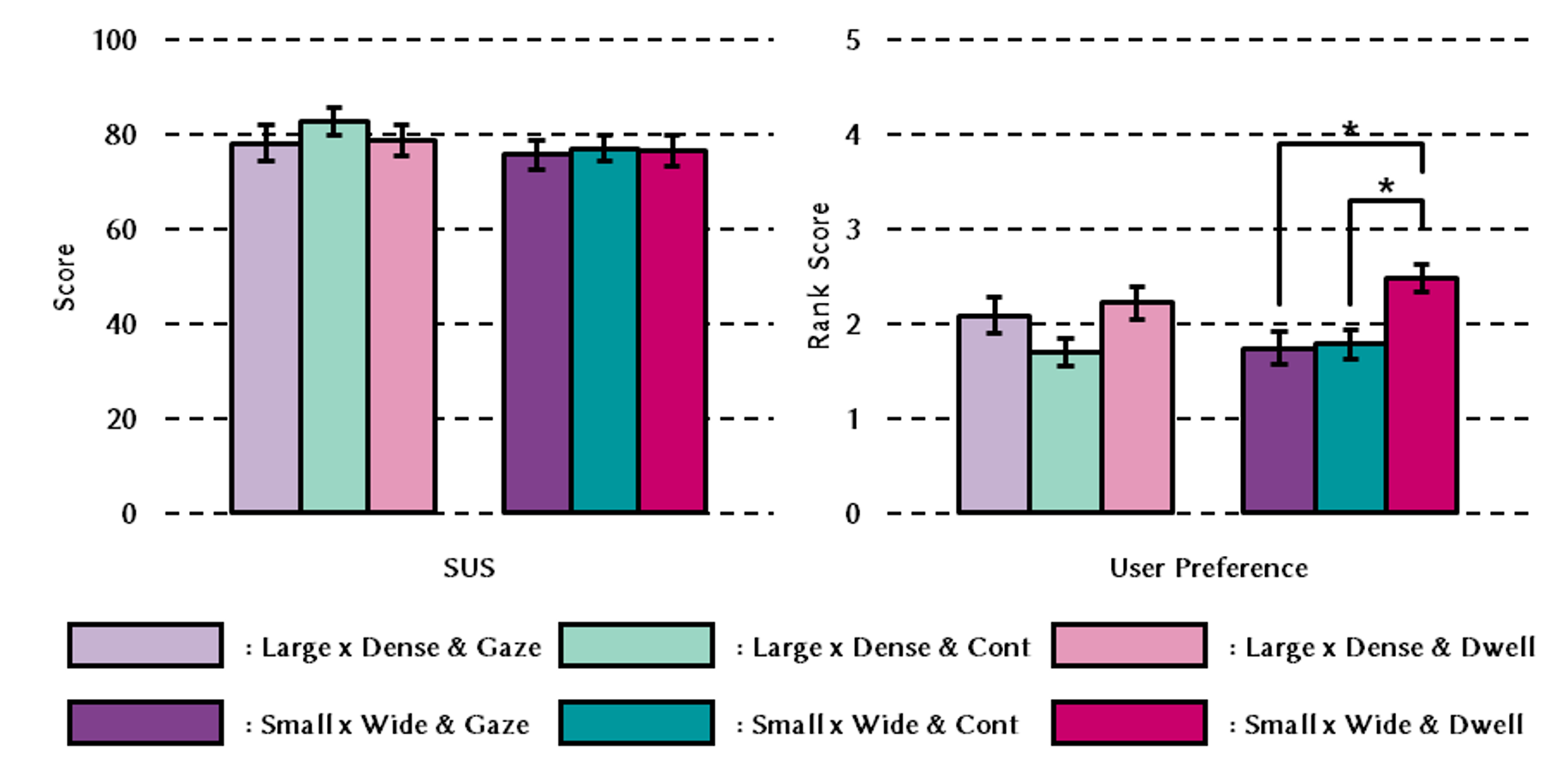}
    \caption{SUS survey score (Left) and user preference rank score collected through interviews (Right) based on target configuration and input modality. Statistically significant differences identified through post-hoc tests are indicated with connecting lines, where *, **, and *** represent p-values less than 0.05, 0.01, and 0.001, respectively.}
    \label{fig:sus}
\end{figure}

The results of the RM-ANOVA analysis indicated that no interaction effect or main effects of Target Configuration and Interaction Technique were observed (see Fig.~\ref{fig:sus} left). Furthermore, no significant differences were found based on Interaction Technique in any of the Target Configurations. This suggests that there were no significant differences in usability between the two factors.

\subsubsection{Interview}

Participants were asked to rank their preference for each interaction technique as 1, 2, or 3 for each configuration, with lower scores indicating higher preference (see Fig.~\ref{fig:sus} right). The results of the Friedman test for the \textit{Large $\times$ Dense} configuration showed no statistically significant differences between the techniques. However, for the \textit{Small $\times$ Wide} configuration, statistically significant differences were found between the techniques. Post-hoc test results revealed that the \textit{Gaze} and \textit{Cont} conditions had smaller ranks score compared to the \textit{Dwell} condition, indicating that participants preferred the \textit{Gaze} and \textit{Cont} conditions in the \textit{Small $\times$ Wide} configuration.

\section{Discussion}

In Study 1, we propose a novel method based on a Bayesian approach to predict selection intention using only gaze data. Furthermore, in Study 2, we demonstrate that accurate and efficient target selection in complex target acquisition environments is achievable by relying solely on predicted selection intention from gaze data. Based on the results of two studies, the contributions of this research can be summarized as follows:

\begin{enumerate}
    \item In Study 1, we confirmed that the proposed Bayesian-based ML model can accurately and quickly discriminate the user's selection intention using only gaze data.
    \item Furthermore, we verified that the approach of converting gaze data into posterior probabilities through a Bayesian method is suitable for using gaze data.
    \item In Study 2, we implemented the interaction technique based on the inference results of our model.
    \item In addition, we demonstrated that our method provides higher accuracy and a more comfortable interaction experience compared to traditional techniques such as controller-based and dwell-based selection.
\end{enumerate}

\subsection{Bayesian Approach to Predicting Selection Intention}

Predicting the user's selection intention is a critical issue in HCI field. By predicting selection intention, it is possible to assist with inaccurate input or provide a fast and convenient interaction experience without the need for manual selection \cite{moon2024:cont:btc:intpre:taskdesign:slc, narkar2024:gaze:dwell:slc:feat:intpre}. Although there have been various attempts to predict selection intention using gaze data \cite{bednarik2012:gaze:intpre, david2021:gaze:feat:slc}, real-time prediction of intention remains a challenging problem due to the noisy nature of gaze data and the need for fast computational processing. Moreover, even when selection intention has been predicted, most approaches have relied on traditional manual selection methods \cite{isomoto2022:gaze:dwell:slc:intpre, narkar2024:gaze:dwell:slc:feat:intpre}.

In this study, inspired by a Bayesian approach that successfully addressed problems in noisy 2D screen touch interactions \cite{bi2013:touch:btc} and 3D target pointing \cite{wei2023:gaze:pt:btc, yu2019:ctrl:pt:btc}, we applied a Bayesian model to convert gaze-related features into posterior probabilities. We hypothesized that these posterior data would be more advantageous for ML models to predict selection intention than traditional observational data formats. Study 1 confirmed this hypothesis. Using the same dataset, prediction performance was significantly higher with posterior data compared to observation data. Furthermore, the prediction performance was comparable to that of previous research, which has reported successful discrimination of selection intention \cite{narkar2024:gaze:dwell:slc:feat:intpre, isomoto2022:gaze:dwell:slc:intpre}.

\subsection{Extending the Use of Gaze Data with Bayesian Approach}

Gaze data can be effectively used as a feature to distinguish or predict an individual's psychological state. It has been widely used in various areas of the HCI field, such as predicting intended targets \cite{wei2023:gaze:pt:btc, david2021:gaze:feat:slc, isomoto2022:gaze:dwell:slc:intpre, narkar2024:gaze:dwell:slc:feat:intpre}, visual attention \cite{jo2024:app:gaze, kim2023:app:gaze}, redirection techniques \cite{jeon2024:app:gaze, sun2018:app}. For example, Jo et al. \cite{jo2024:app:gaze} used gaze-related features to predict most appropriate moments in order to trigger haptic feedback to assist people with peripheral vision loss in a scanning task.Jeon et al. \cite{jeon2024:app:gaze} proposed a redirection technique that uses gaze data to predict the user's trajectory, allowing for longer distance travel with minimal resetting. However, these studies have typically used gaze data directly in the form of observation data. If prediction or classification were attempted using a posterior vector generated through a Bayesian model, we believe the proposed system could provide better performance and an improved user experience.

\subsection{Improving Interaction Experience in 3D Target Selection}

Developing an accurate and efficient 3D target selection technique is a critical issue in VR/AR research, as it supports tasks such as target selection, text input, manipulation, and locomotion. In this study, based on the inference results of the model introduced in Study 1, we propose a novel method that enables target selection using only eye gaze. Through Study 2, we demonstrated that this method provides more accurate and convenient selection compared to the controller-based approach commonly used for target acquisition. Moreover, it provided superior performance over the dwell-based technique, which also uses gaze-based pointing.

\textbf{TTC}: In the results of Study 2, TTC was not affected by the interaction technique. According to previous studies, gaze-based pointing is generally faster than controller-based pointing, which led to the expectation that completion time would vary with the interaction technique \cite{sidenmark2023:gaze:head:ctrl}. However, the experiment results did not support this. This outcome is likely influenced by the complexity of the task used in Study 2. Participants 17 commented, “It took a while to find the right answer for the category word, so there wasn’t much difference in the overall experiment time depending on the technique used.” Since the time taken to locate the correct target dominated over the time required to click the target, effects related to target configuration were observed, while the influence of the interaction technique was not confirmed.

\textbf{Error Rate}: The error rate was higher for controller-based selection than gaze-based selection, which needs the further discussion. The higher rate in the manual selection technique was explained through interviews. Participant 10 said, "I managed to find the target I was looking for, but I kept having trouble with my hand shaking or being unable to point accurately, so I kept trying to click." This outcome aligns with previous studies suggesting that manual selection processes may suffer from reduced selection accuracy due to difficulty in maintaining a steady fixation on the target or movements occurring during the selection process \cite{wu2024:app:acc}. Consequently, manual selection techniques like controller and dwell-based selection showed lower accuracy compared to gaze-based selection, especially in the small and wide target configurations, which highlights this effect.

\subsection{Applications of Interaction Technique without Manual Selection}

In contrast to controller-based selection methods that require both eye and hand coordination, our proposed technique allows for accurate and convenient interaction while keeping both hands free. This characteristic makes it particularly suitable for use as an assistive technology for older adults and people with motor impairments. Wu et al. \cite{wu2024:app:acc} investigated the challenges older adults face when selecting and manipulating 3D objects in VR and found that interaction accuracy decreases during manual selection, leading to negative user experiences and highlighting the need for alternative modalities. The selection technique proposed in this study could address these issues by providing a more accessible solution. In addition, Franz et al.~\cite{franz2023:app:acc} compared different locomotion techniques for individuals with upper limb motor impairments in VR environments and found that methods that did not require a controller were the most preferred. By integrating the proposed selection technique with teleportation, it could serve as a valuable locomotion tool for users with motor impairments.

\subsection{Limitations and Future Work}

To apply the results of this study in real-world settings, several challenges need to be addressed. First, new data must be collected based on the nature of the task and the configuration of the targets to build a robust Bayesian model. This could involve adopting methods such as mechanical simulation-based data collection and neural networks to obtain posterior distributions, as proposed by Moon et al.~\cite{moon2024:cont:btc:intpre:taskdesign:slc, moon2023:btc}. Second, the model developed in Study 1 is currently limited to binary classification. Further research is needed to determine whether multistate classification is feasible for solving problems such as target inference. Finally, because the selection technique was evaluated in an experimental setting with only stationary targets and no distractors, it is crucial to verify its effectiveness in more practical, real-world settings. We hope that our findings will contribute to ongoing research in HCI and VR/AR, especially in addressing fundamental problems using gaze data. Furthermore, we expect that the proposed selection technique will inspire the development of novel interaction methods that provide enhanced user experiences for a wide range of users.

\section{Conclusion}

In VR/AR environments, 3D target acquisition typically relies on interaction techniques that require a manual selection step. These methods often increase the physical and cognitive load, hindering a natural interaction experience. To address this issue, we propose a method that uses gaze data to predict the user's selection intention, enabling interaction without the need for manual selection. Specifically, a Bayesian model is used to transform gaze-related features into posterior data, which is then used to predict user intention through a Bayesian-based ML model. Our model achieved an F1 score of 0.957 and an AUC-ROC performance of 0.988. Furthermore, in Study 2, by comparing our gaze-based selection technique with traditional manual selection methods, the proposed technique demonstrated more accurate selection performance and provided a more comfortable interaction experience. Finally, based on our results, we discuss the advantages of using a Bayesian approach for building interactive systems that use gaze data, and explore how interaction techniques that do not require manual selection can be applied in practical scenarios.

\section{Acknowledgments}
This work was supported by the National Research Foundation of Korea (No. RS-2024-00348094), Korea Radio Promotion Association (No. RNIX20230200) and the Korea Innovation Foundation (No. 2024-IT-RD-0186-01) grant funded by the Korea government (MSIT)

\bibliographystyle{ACM-Reference-Format}
\bibliography{Manuscript}


\begin{thebibliography}{45}


\ifx \showCODEN    \undefined \def \showCODEN     #1{\unskip}     \fi
\ifx \showDOI      \undefined \def \showDOI       #1{#1}\fi
\ifx \showISBNx    \undefined \def \showISBNx     #1{\unskip}     \fi
\ifx \showISBNxiii \undefined \def \showISBNxiii  #1{\unskip}     \fi
\ifx \showISSN     \undefined \def \showISSN      #1{\unskip}     \fi
\ifx \showLCCN     \undefined \def \showLCCN      #1{\unskip}     \fi
\ifx \shownote     \undefined \def \shownote      #1{#1}          \fi
\ifx \showarticletitle \undefined \def \showarticletitle #1{#1}   \fi
\ifx \showURL      \undefined \def \showURL       {\relax}        \fi
\providecommand\bibfield[2]{#2}
\providecommand\bibinfo[2]{#2}
\providecommand\natexlab[1]{#1}
\providecommand\showeprint[2][]{arXiv:#2}

\bibitem[Bednarik et~al\mbox{.}(2012)]%
        {bednarik2012:gaze:intpre}
\bibfield{author}{\bibinfo{person}{Roman Bednarik}, \bibinfo{person}{Hana Vrzakova}, {and} \bibinfo{person}{Michal Hradis}.} \bibinfo{year}{2012}\natexlab{}.
\newblock \showarticletitle{What do you want to do next: a novel approach for intent prediction in gaze-based interaction}. In \bibinfo{booktitle}{\emph{Proceedings of the symposium on eye tracking research and applications}}. \bibinfo{pages}{83--90}.
\newblock


\bibitem[Bi and Zhai(2013)]%
        {bi2013:touch:btc}
\bibfield{author}{\bibinfo{person}{Xiaojun Bi} {and} \bibinfo{person}{Shumin Zhai}.} \bibinfo{year}{2013}\natexlab{}.
\newblock \showarticletitle{Bayesian touch: a statistical criterion of target selection with finger touch}. In \bibinfo{booktitle}{\emph{Proceedings of the 26th annual ACM symposium on User interface software and technology}}. \bibinfo{pages}{51--60}.
\newblock


\bibitem[Boring et~al\mbox{.}(2009)]%
        {boring2009:effort}
\bibfield{author}{\bibinfo{person}{Sebastian Boring}, \bibinfo{person}{Marko Jurmu}, {and} \bibinfo{person}{Andreas Butz}.} \bibinfo{year}{2009}\natexlab{}.
\newblock \showarticletitle{Scroll, tilt or move it: using mobile phones to continuously control pointers on large public displays}. In \bibinfo{booktitle}{\emph{Proceedings of the 21st Annual Conference of the Australian Computer-Human Interaction Special Interest Group: Design: Open 24/7}}. \bibinfo{pages}{161--168}.
\newblock


\bibitem[Bradley(1958)]%
        {bradley1958:latin}
\bibfield{author}{\bibinfo{person}{James~V Bradley}.} \bibinfo{year}{1958}\natexlab{}.
\newblock \showarticletitle{Complete counterbalancing of immediate sequential effects in a Latin square design}.
\newblock \bibinfo{journal}{\emph{J. Amer. Statist. Assoc.}} \bibinfo{volume}{53}, \bibinfo{number}{282} (\bibinfo{year}{1958}), \bibinfo{pages}{525--528}.
\newblock


\bibitem[Brooke(1996)]%
        {brooke1996:sus}
\bibfield{author}{\bibinfo{person}{J Brooke}.} \bibinfo{year}{1996}\natexlab{}.
\newblock \showarticletitle{SUS: A quick and dirty usability scale}.
\newblock \bibinfo{journal}{\emph{Usability Evaluation in Industry}} (\bibinfo{year}{1996}).
\newblock


\bibitem[Choi et~al\mbox{.}(2022)]%
        {choi2022:gaze:head:slc}
\bibfield{author}{\bibinfo{person}{Myungguen Choi}, \bibinfo{person}{Daisuke Sakamoto}, {and} \bibinfo{person}{Tetsuo Ono}.} \bibinfo{year}{2022}\natexlab{}.
\newblock \showarticletitle{Kuiper belt: Utilizing the “out-of-natural angle” region in the eye-gaze interaction for virtual reality}. In \bibinfo{booktitle}{\emph{Proceedings of the 2022 CHI Conference on Human Factors in Computing Systems}}. \bibinfo{pages}{1--17}.
\newblock


\bibitem[David-John et~al\mbox{.}(2021)]%
        {david2021:gaze:feat:slc}
\bibfield{author}{\bibinfo{person}{Brendan David-John}, \bibinfo{person}{Candace Peacock}, \bibinfo{person}{Ting Zhang}, \bibinfo{person}{T~Scott Murdison}, \bibinfo{person}{Hrvoje Benko}, {and} \bibinfo{person}{Tanya~R Jonker}.} \bibinfo{year}{2021}\natexlab{}.
\newblock \showarticletitle{Towards gaze-based prediction of the intent to interact in virtual reality}. In \bibinfo{booktitle}{\emph{ACM Symposium on Eye Tracking Research and Applications}}. \bibinfo{pages}{1--7}.
\newblock


\bibitem[Duchowski(2018)]%
        {duchowski2018:dwell:effort}
\bibfield{author}{\bibinfo{person}{Andrew~T Duchowski}.} \bibinfo{year}{2018}\natexlab{}.
\newblock \showarticletitle{Gaze-based interaction: A 30 year retrospective}.
\newblock \bibinfo{journal}{\emph{Computers \& Graphics}}  \bibinfo{volume}{73} (\bibinfo{year}{2018}), \bibinfo{pages}{59--69}.
\newblock


\bibitem[Duchowski et~al\mbox{.}(2019)]%
        {duchowski2019:gaze:feat:cok}
\bibfield{author}{\bibinfo{person}{Andrew~T Duchowski}, \bibinfo{person}{Krzysztof Krejtz}, \bibinfo{person}{Justyna {\.Z}urawska}, {and} \bibinfo{person}{Donald~H House}.} \bibinfo{year}{2019}\natexlab{}.
\newblock \showarticletitle{Using microsaccades to estimate task difficulty during visual search of layered surfaces}.
\newblock \bibinfo{journal}{\emph{IEEE transactions on visualization and computer graphics}} \bibinfo{volume}{26}, \bibinfo{number}{9} (\bibinfo{year}{2019}), \bibinfo{pages}{2904--2918}.
\newblock


\bibitem[Fernandes and Feiner(2016)]%
        {fernandes2016:survey}
\bibfield{author}{\bibinfo{person}{Ajoy~S Fernandes} {and} \bibinfo{person}{Steven~K Feiner}.} \bibinfo{year}{2016}\natexlab{}.
\newblock \showarticletitle{Combating VR sickness through subtle dynamic field-of-view modification}. In \bibinfo{booktitle}{\emph{2016 IEEE symposium on 3D user interfaces (3DUI)}}. IEEE, \bibinfo{pages}{201--210}.
\newblock


\bibitem[Fitts(1954)]%
        {fitts1954:fitts}
\bibfield{author}{\bibinfo{person}{Paul~M Fitts}.} \bibinfo{year}{1954}\natexlab{}.
\newblock \showarticletitle{The information capacity of the human motor system in controlling the amplitude of movement.}
\newblock \bibinfo{journal}{\emph{Journal of experimental psychology}} \bibinfo{volume}{47}, \bibinfo{number}{6} (\bibinfo{year}{1954}), \bibinfo{pages}{381}.
\newblock


\bibitem[Franz et~al\mbox{.}(2023)]%
        {franz2023:app:acc}
\bibfield{author}{\bibinfo{person}{Rachel~L Franz}, \bibinfo{person}{Jinghan Yu}, {and} \bibinfo{person}{Jacob~O Wobbrock}.} \bibinfo{year}{2023}\natexlab{}.
\newblock \showarticletitle{Comparing Locomotion Techniques in Virtual Reality for People with Upper-Body Motor Impairments}. In \bibinfo{booktitle}{\emph{Proceedings of the 25th International ACM SIGACCESS Conference on Computers and Accessibility}}. \bibinfo{pages}{1--15}.
\newblock


\bibitem[Grossman and Balakrishnan(2005)]%
        {grossman2005:pt:2d}
\bibfield{author}{\bibinfo{person}{Tovi Grossman} {and} \bibinfo{person}{Ravin Balakrishnan}.} \bibinfo{year}{2005}\natexlab{}.
\newblock \showarticletitle{The bubble cursor: enhancing target acquisition by dynamic resizing of the cursor's activation area}. In \bibinfo{booktitle}{\emph{Proceedings of the SIGCHI conference on Human factors in computing systems}}. \bibinfo{pages}{281--290}.
\newblock


\bibitem[Hart(1988)]%
        {hart1988:nasa}
\bibfield{author}{\bibinfo{person}{SG Hart}.} \bibinfo{year}{1988}\natexlab{}.
\newblock \showarticletitle{Development of NASA-TLX (Task Load Index): Results of empirical and theoretical research}.
\newblock \bibinfo{journal}{\emph{Human mental workload/Elsevier}} (\bibinfo{year}{1988}).
\newblock


\bibitem[Hincapi{\'e}-Ramos et~al\mbox{.}(2014)]%
        {hincapie2014:hand:ctrl:effort}
\bibfield{author}{\bibinfo{person}{Juan~David Hincapi{\'e}-Ramos}, \bibinfo{person}{Xiang Guo}, \bibinfo{person}{Paymahn Moghadasian}, {and} \bibinfo{person}{Pourang Irani}.} \bibinfo{year}{2014}\natexlab{}.
\newblock \showarticletitle{Consumed endurance: a metric to quantify arm fatigue of mid-air interactions}. In \bibinfo{booktitle}{\emph{Proceedings of the SIGCHI Conference on Human Factors in Computing Systems}}. \bibinfo{pages}{1063--1072}.
\newblock


\bibitem[Isomoto et~al\mbox{.}(2018)]%
        {isomoto2018:gaze:dwell:slc:intpre}
\bibfield{author}{\bibinfo{person}{Toshiya Isomoto}, \bibinfo{person}{Toshiyuki Ando}, \bibinfo{person}{Buntarou Shizuki}, {and} \bibinfo{person}{Shin Takahashi}.} \bibinfo{year}{2018}\natexlab{}.
\newblock \showarticletitle{Dwell time reduction technique using Fitts' law for gaze-based target acquisition}. In \bibinfo{booktitle}{\emph{Proceedings of the 2018 ACM Symposium on Eye Tracking Research \& Applications}}. \bibinfo{pages}{1--7}.
\newblock


\bibitem[Isomoto et~al\mbox{.}(2022)]%
        {isomoto2022:gaze:dwell:slc:intpre}
\bibfield{author}{\bibinfo{person}{Toshiya Isomoto}, \bibinfo{person}{Shota Yamanaka}, {and} \bibinfo{person}{Buntarou Shizuki}.} \bibinfo{year}{2022}\natexlab{}.
\newblock \showarticletitle{Dwell selection with ML-based intent prediction using only gaze data}.
\newblock \bibinfo{journal}{\emph{Proceedings of the ACM on Interactive, Mobile, Wearable and Ubiquitous Technologies}} \bibinfo{volume}{6}, \bibinfo{number}{3} (\bibinfo{year}{2022}), \bibinfo{pages}{1--21}.
\newblock


\bibitem[Jacob(1990a)]%
        {jacob1990:gaze:midas}
\bibfield{author}{\bibinfo{person}{Robert~JK Jacob}.} \bibinfo{year}{1990}\natexlab{a}.
\newblock \showarticletitle{What you look at is what you get: eye movement-based interaction techniques}. In \bibinfo{booktitle}{\emph{Proceedings of the SIGCHI conference on Human factors in computing systems}}. \bibinfo{pages}{11--18}.
\newblock


\bibitem[Jacob(1990b)]%
        {jacob1990:gaze:dwell:slc}
\bibfield{author}{\bibinfo{person}{Robert~JK Jacob}.} \bibinfo{year}{1990}\natexlab{b}.
\newblock \showarticletitle{What you look at is what you get: eye movement-based interaction techniques}. In \bibinfo{booktitle}{\emph{Proceedings of the SIGCHI conference on Human factors in computing systems}}. \bibinfo{pages}{11--18}.
\newblock


\bibitem[Jeon et~al\mbox{.}(2024)]%
        {jeon2024:app:gaze}
\bibfield{author}{\bibinfo{person}{Sang-Bin Jeon}, \bibinfo{person}{Jaeho Jung}, \bibinfo{person}{Jinhyung Park}, {and} \bibinfo{person}{In-Kwon Lee}.} \bibinfo{year}{2024}\natexlab{}.
\newblock \showarticletitle{F-RDW: Redirected Walking with Forecasting Future Position}.
\newblock \bibinfo{journal}{\emph{IEEE Transactions on Visualization and Computer Graphics}} (\bibinfo{year}{2024}).
\newblock


\bibitem[Jo et~al\mbox{.}(2024)]%
        {jo2024:app:gaze}
\bibfield{author}{\bibinfo{person}{Taewoo Jo}, \bibinfo{person}{Dohyeon Yeo}, \bibinfo{person}{Gwangbin Kim}, \bibinfo{person}{Seokhyun Hwang}, {and} \bibinfo{person}{SeungJun Kim}.} \bibinfo{year}{2024}\natexlab{}.
\newblock \showarticletitle{WatchCap: Improving Scanning Efficiency in People with Low Vision through Compensatory Head Movement Stimulation}.
\newblock \bibinfo{journal}{\emph{Proceedings of the ACM on Interactive, Mobile, Wearable and Ubiquitous Technologies}} \bibinfo{volume}{8}, \bibinfo{number}{2} (\bibinfo{year}{2024}), \bibinfo{pages}{1--32}.
\newblock


\bibitem[Kim et~al\mbox{.}(2023)]%
        {kim2023:app:gaze}
\bibfield{author}{\bibinfo{person}{Gwangbin Kim}, \bibinfo{person}{Dohyeon Yeo}, \bibinfo{person}{Taewoo Jo}, \bibinfo{person}{Daniela Rus}, {and} \bibinfo{person}{SeungJun Kim}.} \bibinfo{year}{2023}\natexlab{}.
\newblock \showarticletitle{What and When to Explain? On-road Evaluation of Explanations in Highly Automated Vehicles}.
\newblock \bibinfo{journal}{\emph{Proceedings of the ACM on Interactive, Mobile, Wearable and Ubiquitous Technologies}} \bibinfo{volume}{7}, \bibinfo{number}{3} (\bibinfo{year}{2023}), \bibinfo{pages}{1--26}.
\newblock


\bibitem[Krejtz et~al\mbox{.}(2016)]%
        {krejtz2016:cok}
\bibfield{author}{\bibinfo{person}{Krzysztof Krejtz}, \bibinfo{person}{Andrew Duchowski}, \bibinfo{person}{Izabela Krejtz}, \bibinfo{person}{Agnieszka Szarkowska}, {and} \bibinfo{person}{Agata Kopacz}.} \bibinfo{year}{2016}\natexlab{}.
\newblock \showarticletitle{Discerning ambient/focal attention with coefficient K}.
\newblock \bibinfo{journal}{\emph{ACM Transactions on Applied Perception (TAP)}} \bibinfo{volume}{13}, \bibinfo{number}{3} (\bibinfo{year}{2016}), \bibinfo{pages}{1--20}.
\newblock


\bibitem[Li et~al\mbox{.}(2021)]%
        {li2021:gaze:2d:btc}
\bibfield{author}{\bibinfo{person}{Zhi Li}, \bibinfo{person}{Maozheng Zhao}, \bibinfo{person}{Yifan Wang}, \bibinfo{person}{Sina Rashidian}, \bibinfo{person}{Furqan Baig}, \bibinfo{person}{Rui Liu}, \bibinfo{person}{Wanyu Liu}, \bibinfo{person}{Michel Beaudouin-Lafon}, \bibinfo{person}{Brooke Ellison}, \bibinfo{person}{Fusheng Wang}, {et~al\mbox{.}}} \bibinfo{year}{2021}\natexlab{}.
\newblock \showarticletitle{Bayesgaze: A bayesian approach to eye-gaze based target selection}. In \bibinfo{booktitle}{\emph{Proceedings. Graphics Interface (Conference)}}, Vol.~\bibinfo{volume}{2021}. NIH Public Access, \bibinfo{pages}{231}.
\newblock


\bibitem[Lu et~al\mbox{.}(2021)]%
        {lu2021:blink}
\bibfield{author}{\bibinfo{person}{Xueshi Lu}, \bibinfo{person}{Difeng Yu}, \bibinfo{person}{Hai-Ning Liang}, {and} \bibinfo{person}{Jorge Goncalves}.} \bibinfo{year}{2021}\natexlab{}.
\newblock \showarticletitle{itext: Hands-free text entry on an imaginary keyboard for augmented reality systems}. In \bibinfo{booktitle}{\emph{The 34th Annual ACM Symposium on User Interface Software and Technology}}. \bibinfo{pages}{815--825}.
\newblock


\bibitem[Lu et~al\mbox{.}(2020)]%
        {lu2020:cont:slc:taskdesign}
\bibfield{author}{\bibinfo{person}{Yiqin Lu}, \bibinfo{person}{Chun Yu}, {and} \bibinfo{person}{Yuanchun Shi}.} \bibinfo{year}{2020}\natexlab{}.
\newblock \showarticletitle{Investigating bubble mechanism for ray-casting to improve 3d target acquisition in virtual reality}. In \bibinfo{booktitle}{\emph{2020 IEEE Conference on virtual reality and 3D user interfaces (VR)}}. IEEE, \bibinfo{pages}{35--43}.
\newblock


\bibitem[MacKenzie(1992)]%
        {mackenzie1992:fitts}
\bibfield{author}{\bibinfo{person}{I~Scott MacKenzie}.} \bibinfo{year}{1992}\natexlab{}.
\newblock \showarticletitle{Fitts' law as a research and design tool in human-computer interaction}.
\newblock \bibinfo{journal}{\emph{Human-computer interaction}} \bibinfo{volume}{7}, \bibinfo{number}{1} (\bibinfo{year}{1992}), \bibinfo{pages}{91--139}.
\newblock


\bibitem[Moon et~al\mbox{.}(2024)]%
        {moon2024:cont:btc:intpre:taskdesign:slc}
\bibfield{author}{\bibinfo{person}{Hee-Seung Moon}, \bibinfo{person}{Yi-Chi Liao}, \bibinfo{person}{Chenyu Li}, \bibinfo{person}{Byungjoo Lee}, {and} \bibinfo{person}{Antti Oulasvirta}.} \bibinfo{year}{2024}\natexlab{}.
\newblock \showarticletitle{Real-time 3D Target Inference via Biomechanical Simulation}. In \bibinfo{booktitle}{\emph{Proceedings of the CHI Conference on Human Factors in Computing Systems}}. \bibinfo{pages}{1--18}.
\newblock


\bibitem[Moon et~al\mbox{.}(2023)]%
        {moon2023:btc}
\bibfield{author}{\bibinfo{person}{Hee-Seung Moon}, \bibinfo{person}{Antti Oulasvirta}, {and} \bibinfo{person}{Byungjoo Lee}.} \bibinfo{year}{2023}\natexlab{}.
\newblock \showarticletitle{Amortized inference with user simulations}. In \bibinfo{booktitle}{\emph{Proceedings of the 2023 CHI Conference on Human Factors in Computing Systems}}. \bibinfo{pages}{1--20}.
\newblock


\bibitem[Narkar et~al\mbox{.}(2024)]%
        {narkar2024:gaze:dwell:slc:feat:intpre}
\bibfield{author}{\bibinfo{person}{Anish~S Narkar}, \bibinfo{person}{Jan~J Michalak}, \bibinfo{person}{Candace~E Peacock}, {and} \bibinfo{person}{Brendan David-John}.} \bibinfo{year}{2024}\natexlab{}.
\newblock \showarticletitle{GazeIntent: Adapting dwell-time selection in VR interaction with real-time intent modeling}.
\newblock \bibinfo{journal}{\emph{arXiv preprint arXiv:2404.13829}} (\bibinfo{year}{2024}).
\newblock


\bibitem[Periverzov and Ilie{\c{s}}(2015)]%
        {periverzov2015:hand:slc}
\bibfield{author}{\bibinfo{person}{Frol Periverzov} {and} \bibinfo{person}{Horea Ilie{\c{s}}}.} \bibinfo{year}{2015}\natexlab{}.
\newblock \showarticletitle{IDS: The intent driven selection method for natural user interfaces}. In \bibinfo{booktitle}{\emph{2015 IEEE symposium on 3D user interfaces (3DUI)}}. IEEE, \bibinfo{pages}{121--128}.
\newblock


\bibitem[Ren et~al\mbox{.}(2024)]%
        {ren2024:gaze:hand:btc}
\bibfield{author}{\bibinfo{person}{Yunlei Ren}, \bibinfo{person}{Yan Zhang}, \bibinfo{person}{Zhitao Liu}, \bibinfo{person}{Yi Li}, \bibinfo{person}{Li Yuan}, {and} \bibinfo{person}{Ning Xie}.} \bibinfo{year}{2024}\natexlab{}.
\newblock \showarticletitle{Eye-Hand Typing: Eye Gaze Assisted Finger Typing via Bayesian Processes in AR}.
\newblock \bibinfo{journal}{\emph{IEEE Transactions on Visualization and Computer Graphics}} (\bibinfo{year}{2024}).
\newblock


\bibitem[Sendhilnathan et~al\mbox{.}(2022)]%
        {sendhilnathan2022:ivdt:error}
\bibfield{author}{\bibinfo{person}{Naveen Sendhilnathan}, \bibinfo{person}{Ting Zhang}, \bibinfo{person}{Ben Lafreniere}, \bibinfo{person}{Tovi Grossman}, {and} \bibinfo{person}{Tanya~R Jonker}.} \bibinfo{year}{2022}\natexlab{}.
\newblock \showarticletitle{Detecting input recognition errors and user errors using gaze dynamics in virtual reality}. In \bibinfo{booktitle}{\emph{Proceedings of the 35th Annual ACM Symposium on User Interface Software and Technology}}. \bibinfo{pages}{1--19}.
\newblock


\bibitem[Shi et~al\mbox{.}(2023)]%
        {shi2023:gaze:hand:slc}
\bibfield{author}{\bibinfo{person}{Rongkai Shi}, \bibinfo{person}{Yushi Wei}, \bibinfo{person}{Xueying Qin}, \bibinfo{person}{Pan Hui}, {and} \bibinfo{person}{Hai-Ning Liang}.} \bibinfo{year}{2023}\natexlab{}.
\newblock \showarticletitle{Exploring gaze-assisted and hand-based region selection in augmented reality}.
\newblock \bibinfo{journal}{\emph{Proceedings of the ACM on Human-Computer Interaction}} \bibinfo{volume}{7}, \bibinfo{number}{ETRA} (\bibinfo{year}{2023}), \bibinfo{pages}{1--19}.
\newblock


\bibitem[Sidenmark et~al\mbox{.}(2023)]%
        {sidenmark2023:gaze:head:ctrl}
\bibfield{author}{\bibinfo{person}{Ludwig Sidenmark}, \bibinfo{person}{Franziska Prummer}, \bibinfo{person}{Joshua Newn}, {and} \bibinfo{person}{Hans Gellersen}.} \bibinfo{year}{2023}\natexlab{}.
\newblock \showarticletitle{Comparing Gaze, Head and Controller Selection of Dynamically Revealed Targets in Head-mounted Displays}.
\newblock \bibinfo{journal}{\emph{IEEE Transactions on Visualization and Computer Graphics}} (\bibinfo{year}{2023}).
\newblock


\bibitem[Sun et~al\mbox{.}(2018)]%
        {sun2018:app}
\bibfield{author}{\bibinfo{person}{Qi Sun}, \bibinfo{person}{Anjul Patney}, \bibinfo{person}{Li-Yi Wei}, \bibinfo{person}{Omer Shapira}, \bibinfo{person}{Jingwan Lu}, \bibinfo{person}{Paul Asente}, \bibinfo{person}{Suwen Zhu}, \bibinfo{person}{Morgan McGuire}, \bibinfo{person}{David Luebke}, {and} \bibinfo{person}{Arie Kaufman}.} \bibinfo{year}{2018}\natexlab{}.
\newblock \showarticletitle{Towards virtual reality infinite walking: dynamic saccadic redirection}.
\newblock \bibinfo{journal}{\emph{ACM Transactions on Graphics (TOG)}} \bibinfo{volume}{37}, \bibinfo{number}{4} (\bibinfo{year}{2018}), \bibinfo{pages}{1--13}.
\newblock


\bibitem[Tanriverdi and Jacob(2000)]%
        {tanriverdi2000:gaze:hand}
\bibfield{author}{\bibinfo{person}{Vildan Tanriverdi} {and} \bibinfo{person}{Robert~JK Jacob}.} \bibinfo{year}{2000}\natexlab{}.
\newblock \showarticletitle{Interacting with eye movements in virtual environments}. In \bibinfo{booktitle}{\emph{Proceedings of the SIGCHI conference on Human Factors in Computing Systems}}. \bibinfo{pages}{265--272}.
\newblock


\bibitem[Vertegaal(2008)]%
        {vertegaal2008:gaze}
\bibfield{author}{\bibinfo{person}{Roel Vertegaal}.} \bibinfo{year}{2008}\natexlab{}.
\newblock \showarticletitle{A Fitts Law comparison of eye tracking and manual input in the selection of visual targets}. In \bibinfo{booktitle}{\emph{Proceedings of the 10th international conference on Multimodal interfaces}}. \bibinfo{pages}{241--248}.
\newblock


\bibitem[Vidal et~al\mbox{.}(2013)]%
        {vidal2013:pursuits}
\bibfield{author}{\bibinfo{person}{M{\'e}lodie Vidal}, \bibinfo{person}{Andreas Bulling}, {and} \bibinfo{person}{Hans Gellersen}.} \bibinfo{year}{2013}\natexlab{}.
\newblock \showarticletitle{Pursuits: spontaneous interaction with displays based on smooth pursuit eye movement and moving targets}. In \bibinfo{booktitle}{\emph{Proceedings of the 2013 ACM international joint conference on Pervasive and ubiquitous computing}}. \bibinfo{pages}{439--448}.
\newblock


\bibitem[Wei et~al\mbox{.}(2023)]%
        {wei2023:gaze:pt:btc}
\bibfield{author}{\bibinfo{person}{Yushi Wei}, \bibinfo{person}{Rongkai Shi}, \bibinfo{person}{Difeng Yu}, \bibinfo{person}{Yihong Wang}, \bibinfo{person}{Yue Li}, \bibinfo{person}{Lingyun Yu}, {and} \bibinfo{person}{Hai-Ning Liang}.} \bibinfo{year}{2023}\natexlab{}.
\newblock \showarticletitle{Predicting gaze-based target selection in augmented reality headsets based on eye and head endpoint distributions}. In \bibinfo{booktitle}{\emph{Proceedings of the 2023 CHI Conference on Human Factors in Computing Systems}}. \bibinfo{pages}{1--14}.
\newblock


\bibitem[Wu et~al\mbox{.}(2024)]%
        {wu2024:app:acc}
\bibfield{author}{\bibinfo{person}{Zhiqing Wu}, \bibinfo{person}{Duotun Wang}, \bibinfo{person}{Shumeng Zhang}, \bibinfo{person}{Yuru Huang}, \bibinfo{person}{Zeyu Wang}, {and} \bibinfo{person}{Mingming Fan}.} \bibinfo{year}{2024}\natexlab{}.
\newblock \showarticletitle{Toward Making Virtual Reality (VR) More Inclusive for Older Adults: Investigating Aging Effect on Target Selection and Manipulation Tasks in VR}. In \bibinfo{booktitle}{\emph{Proceedings of the CHI Conference on Human Factors in Computing Systems}}. \bibinfo{pages}{1--17}.
\newblock


\bibitem[Yi et~al\mbox{.}(2022)]%
        {yi2022:gaze:slc:eyelid}
\bibfield{author}{\bibinfo{person}{Xin Yi}, \bibinfo{person}{Leping Qiu}, \bibinfo{person}{Wenjing Tang}, \bibinfo{person}{Yehan Fan}, \bibinfo{person}{Hewu Li}, {and} \bibinfo{person}{Yuanchun Shi}.} \bibinfo{year}{2022}\natexlab{}.
\newblock \showarticletitle{Deep: 3d gaze pointing in virtual reality leveraging eyelid movement}. In \bibinfo{booktitle}{\emph{Proceedings of the 35th Annual ACM Symposium on User Interface Software and Technology}}. \bibinfo{pages}{1--14}.
\newblock


\bibitem[Yu et~al\mbox{.}(2019)]%
        {yu2019:ctrl:pt:btc}
\bibfield{author}{\bibinfo{person}{Difeng Yu}, \bibinfo{person}{Hai-Ning Liang}, \bibinfo{person}{Xueshi Lu}, \bibinfo{person}{Kaixuan Fan}, {and} \bibinfo{person}{Barrett Ens}.} \bibinfo{year}{2019}\natexlab{}.
\newblock \showarticletitle{Modeling endpoint distribution of pointing selection tasks in virtual reality environments}.
\newblock \bibinfo{journal}{\emph{ACM Transactions on Graphics (TOG)}} \bibinfo{volume}{38}, \bibinfo{number}{6} (\bibinfo{year}{2019}), \bibinfo{pages}{1--13}.
\newblock


\bibitem[Zhai and Kristensson(2012)]%
        {zhai2012:btc}
\bibfield{author}{\bibinfo{person}{Shumin Zhai} {and} \bibinfo{person}{Per~Ola Kristensson}.} \bibinfo{year}{2012}\natexlab{}.
\newblock \showarticletitle{The word-gesture keyboard: reimagining keyboard interaction}.
\newblock \bibinfo{journal}{\emph{Commun. ACM}} \bibinfo{volume}{55}, \bibinfo{number}{9} (\bibinfo{year}{2012}), \bibinfo{pages}{91--101}.
\newblock


\bibitem[Zhu et~al\mbox{.}(2023)]%
        {zhu2023:hand:slc:pinch}
\bibfield{author}{\bibinfo{person}{Fengyuan Zhu}, \bibinfo{person}{Ludwig Sidenmark}, \bibinfo{person}{Mauricio Sousa}, {and} \bibinfo{person}{Tovi Grossman}.} \bibinfo{year}{2023}\natexlab{}.
\newblock \showarticletitle{PinchLens: Applying Spatial Magnification and Adaptive Control-Display Gain for Precise Selection in Virtual Reality}. In \bibinfo{booktitle}{\emph{2023 IEEE International Symposium on Mixed and Augmented Reality (ISMAR)}}. IEEE, \bibinfo{pages}{1221--1230}.
\newblock


\end{thebibliography}

\end{document}


\title{Supplementary Material: \newline"Predicting Selection Intention in Real-Time with Bayesian-based ML Model in Unimodal Gaze Interaction"}

\maketitle

\appendix

\section{Bayesian Model}

This section discusses the detailed information of the Bayesian-based ML model constructed in Study 1 and utilized in Study 2. The Bayesian-based ML model predicts the user's selection intention by utilizing gaze-related features calculated through the gaze data pipeline. Specifically, selected gaze-related features, chosen as observation data, are converted into a posterior vector, which is then passed to the ML model to classify the selection intention. The following section provides the rationale for selecting the features as observation data and presents the results of fitting these features to a probability density function (PDF). Additionally, this section includes details on the type of ML model used to classify selection intention, as well as the dataset, hyperparameters, and other relevant information employed during the training process.

\subsection{Feature Selection}

\begin{table}[ht]
\caption{The results of the independent t-test performed on the features from the 1st and 2nd Fixation groups, based on class type}
\label{tab:metric:selection:part1}
\resizebox{\textwidth}{!}{%
\begin{tabular}{cccllllllll}
\hline
\multirow{3}{*}{\textbf{\begin{tabular}[c]{@{}c@{}}Task \\ Type\end{tabular}}} &
  \multirow{3}{*}{\textbf{\begin{tabular}[c]{@{}c@{}}Target \\ Width\end{tabular}}} &
  \multirow{3}{*}{\textbf{\begin{tabular}[c]{@{}c@{}}Target \\ Density\end{tabular}}} &
  \multicolumn{8}{c}{\textbf{Gaze-realted Features}} \\ \cline{4-11} 
 &
   &
   &
  \multicolumn{4}{c}{\textbf{1st Fixation}} &
  \multicolumn{4}{c}{\textbf{2nd Fixation}} \\ \cline{4-11} 
 &
   &
   &
  \multicolumn{1}{c}{\begin{tabular}[c]{@{}c@{}}Fixation\\ duration\end{tabular}} &
  \multicolumn{1}{c}{\begin{tabular}[c]{@{}c@{}}Fixation\\ std x\end{tabular}} &
  \multicolumn{1}{c}{\begin{tabular}[c]{@{}c@{}}Fixation\\ std y\end{tabular}} &
  \multicolumn{1}{c}{\begin{tabular}[c]{@{}c@{}}Fixation\\ velocity\end{tabular}} &
  \multicolumn{1}{c}{\begin{tabular}[c]{@{}c@{}}Fixation\\ duration\end{tabular}} &
  \multicolumn{1}{c}{\begin{tabular}[c]{@{}c@{}}Fixation\\ std x\end{tabular}} &
  \multicolumn{1}{c}{\begin{tabular}[c]{@{}c@{}}Fixation\\ std y\end{tabular}} &
  \multicolumn{1}{c}{\begin{tabular}[c]{@{}c@{}}Fixation\\ velocity\end{tabular}} \\ \hline
\textit{\textbf{Simple}} &
  \textit{\textbf{Large}} &
  \textit{\textbf{Dense}} &
  \begin{tabular}[c]{@{}l@{}}p < .001\\ d = 0.314\end{tabular} &
  \begin{tabular}[c]{@{}l@{}}p < .001\\ d = 0.418\end{tabular} &
  \begin{tabular}[c]{@{}l@{}}p < .001\\ d = 0.432\end{tabular} &
  \begin{tabular}[c]{@{}l@{}}p < .001\\ d = 0.353\end{tabular} &
  \begin{tabular}[c]{@{}l@{}}p < .001\\ d = -1.651\end{tabular} &
  \begin{tabular}[c]{@{}l@{}}p < .001\\ d = -0.487\end{tabular} &
  \begin{tabular}[c]{@{}l@{}}p < .001\\ d = -0.62\end{tabular} &
  \begin{tabular}[c]{@{}l@{}}p < .001\\ d = 0.238\end{tabular} \\
\textit{\textbf{Simple}} &
  \textit{\textbf{Large}} &
  \textit{\textbf{Wide}} &
  \begin{tabular}[c]{@{}l@{}}p = 0.071\\ d = -0.081\end{tabular} &
  \begin{tabular}[c]{@{}l@{}}p = 0.003\\ d = -0.131\end{tabular} &
  \begin{tabular}[c]{@{}l@{}}p = 0.005\\ d = -0.124\end{tabular} &
  \begin{tabular}[c]{@{}l@{}}p < .001\\ d = -0.172\end{tabular} &
  \begin{tabular}[c]{@{}l@{}}p < .001\\ d = -1.261\end{tabular} &
  \begin{tabular}[c]{@{}l@{}}p < .001\\ d = -0.66\end{tabular} &
  \begin{tabular}[c]{@{}l@{}}p < .001\\ d = -0.677\end{tabular} &
  \begin{tabular}[c]{@{}l@{}}p = 0.034\\ d = 0.086\end{tabular} \\
\textit{\textbf{Simple}} &
  \textit{\textbf{Small}} &
  \textit{\textbf{Dense}} &
  \begin{tabular}[c]{@{}l@{}}p < .001\\ d = 0.576\end{tabular} &
  \begin{tabular}[c]{@{}l@{}}p < .001\\ d = 0.351\end{tabular} &
  \begin{tabular}[c]{@{}l@{}}p < .001\\ d = 0.231\end{tabular} &
  \begin{tabular}[c]{@{}l@{}}p < .001\\ d = 0.23\end{tabular} &
  \begin{tabular}[c]{@{}l@{}}p < .001\\ d = -1.742\end{tabular} &
  \begin{tabular}[c]{@{}l@{}}p < .001\\ d = -0.786\end{tabular} &
  \begin{tabular}[c]{@{}l@{}}p < .001\\ d = -0.901\end{tabular} &
  \begin{tabular}[c]{@{}l@{}}p < .001\\ d = 0.255\end{tabular} \\
\textit{\textbf{Simple}} &
  \textit{\textbf{Small}} &
  \textit{\textbf{Wide}} &
  \begin{tabular}[c]{@{}l@{}}p = 0.537\\ d = -0.025\end{tabular} &
  \begin{tabular}[c]{@{}l@{}}p < .001\\ d = -0.323\end{tabular} &
  \begin{tabular}[c]{@{}l@{}}p < .001\\ d = -0.176\end{tabular} &
  \begin{tabular}[c]{@{}l@{}}p < .001\\ d = -0.337\end{tabular} &
  \begin{tabular}[c]{@{}l@{}}p < .001\\ d = -1.441\end{tabular} &
  \begin{tabular}[c]{@{}l@{}}p < .001\\ d = -0.805\end{tabular} &
  \begin{tabular}[c]{@{}l@{}}p < .001\\ d = -0.728\end{tabular} &
  \begin{tabular}[c]{@{}l@{}}p < .001\\ d = 0.306\end{tabular} \\
\textit{\textbf{Complex}} &
  \textit{\textbf{Large}} &
  \textit{\textbf{Dense}} &
  \begin{tabular}[c]{@{}l@{}}p = 0.101\\ d = 0.064\end{tabular} &
  \begin{tabular}[c]{@{}l@{}}p = 0.023\\ d = 0.09\end{tabular} &
  \begin{tabular}[c]{@{}l@{}}p = 0.103\\ d = 0.065\end{tabular} &
  \begin{tabular}[c]{@{}l@{}}p < .001\\ d = 0.159\end{tabular} &
  \begin{tabular}[c]{@{}l@{}}p < .001\\ d = -0.723\end{tabular} &
  \begin{tabular}[c]{@{}l@{}}p < .001\\ d = -0.319\end{tabular} &
  \begin{tabular}[c]{@{}l@{}}p < .001\\ d = -0.495\end{tabular} &
  \begin{tabular}[c]{@{}l@{}}p = 0.006\\ d = 0.101\end{tabular} \\
\textit{\textbf{Complex}} &
  \textit{\textbf{Large}} &
  \textit{\textbf{Wide}} &
  \begin{tabular}[c]{@{}l@{}}p = 0.129\\ d = -0.059\end{tabular} &
  \begin{tabular}[c]{@{}l@{}}p < .001\\ d = 0.154\end{tabular} &
  \begin{tabular}[c]{@{}l@{}}p < .001\\ d = 0.175\end{tabular} &
  \begin{tabular}[c]{@{}l@{}}p < .001\\ d = 0.6\end{tabular} &
  \begin{tabular}[c]{@{}l@{}}p < .001\\ d = -0.754\end{tabular} &
  \begin{tabular}[c]{@{}l@{}}p < .001\\ d = -0.273\end{tabular} &
  \begin{tabular}[c]{@{}l@{}}p < .001\\ d = -0.331\end{tabular} &
  \begin{tabular}[c]{@{}l@{}}p < .001\\ d = 0.677\end{tabular} \\
\textit{\textbf{Complex}} &
  \textit{\textbf{Small}} &
  \textit{\textbf{Dense}} &
  \begin{tabular}[c]{@{}l@{}}p < .001\\ d = 0.213\end{tabular} &
  \begin{tabular}[c]{@{}l@{}}p = 0.002\\ d = 0.114\end{tabular} &
  \begin{tabular}[c]{@{}l@{}}p = 0.001\\ d = 0.117\end{tabular} &
  \begin{tabular}[c]{@{}l@{}}p < .001\\ d = 0.16\end{tabular} &
  \begin{tabular}[c]{@{}l@{}}p < .001\\ d = -0.895\end{tabular} &
  \begin{tabular}[c]{@{}l@{}}p < .001\\ d = -0.486\end{tabular} &
  \begin{tabular}[c]{@{}l@{}}p < .001\\ d = -0.588\end{tabular} &
  \begin{tabular}[c]{@{}l@{}}p = 0.002\\ d = 0.099\end{tabular} \\
\textit{\textbf{Complex}} &
  \textit{\textbf{Small}} &
  \textit{\textbf{Wide}} &
  \begin{tabular}[c]{@{}l@{}}p < .001\\ d = 0.195\end{tabular} &
  \begin{tabular}[c]{@{}l@{}}p < .001\\ d = 0.207\end{tabular} &
  \begin{tabular}[c]{@{}l@{}}p < .001\\ d = 0.221\end{tabular} &
  \begin{tabular}[c]{@{}l@{}}p < .001\\ d = 0.468\end{tabular} &
  \begin{tabular}[c]{@{}l@{}}p < .001\\ d = -1.007\end{tabular} &
  \begin{tabular}[c]{@{}l@{}}p < .001\\ d = -0.373\end{tabular} &
  \begin{tabular}[c]{@{}l@{}}p < .001\\ d = -0.483\end{tabular} &
  \begin{tabular}[c]{@{}l@{}}p < .001\\ d = 0.703\end{tabular} \\ \hline
\multicolumn{3}{c}{\textbf{Total Count}} &
  4 &
  \textbf{8} &
  \textbf{7} &
  \textbf{8} &
  \textbf{8} &
  \textbf{8} &
  \textbf{8} &
  \textbf{8} \\ \hline
\end{tabular}%
}
\end{table}

\begin{table}[ht]
\caption{The results of the independent t-test performed on the features from the Saccade and Coefficient K groups, based on class type}
\label{tab:metric:selection:part2}
\resizebox{\textwidth}{!}{%
\begin{tabular}{cccllllllll}
\hline
\multirow{3}{*}{\textbf{\begin{tabular}[c]{@{}c@{}}Task \\ Type\end{tabular}}} &
  \multirow{3}{*}{\textbf{\begin{tabular}[c]{@{}c@{}}Target \\ Width\end{tabular}}} &
  \multirow{3}{*}{\textbf{\begin{tabular}[c]{@{}c@{}}Target \\ Density\end{tabular}}} &
  \multicolumn{8}{c}{\textbf{Gaze-related Features}} \\ \cline{4-11} 
 &
   &
   &
  \multicolumn{7}{c}{\textbf{Saccade}} &
  \multicolumn{1}{c}{\textbf{Coefficient K}} \\ \cline{4-11} 
 &
   &
   &
  \multicolumn{1}{c}{\begin{tabular}[c]{@{}c@{}}Saccade\\ duration\end{tabular}} &
  \multicolumn{1}{c}{\begin{tabular}[c]{@{}c@{}}Saccade\\ amplitude\\ eye\end{tabular}} &
  \multicolumn{1}{c}{\begin{tabular}[c]{@{}c@{}}Saccade\\ amplitude\\ head\end{tabular}} &
  \multicolumn{1}{c}{\begin{tabular}[c]{@{}c@{}}Saccade\\ amplitude\\ gaze\end{tabular}} &
  \multicolumn{1}{c}{\begin{tabular}[c]{@{}c@{}}Saccade\\ velocity\\ eye\end{tabular}} &
  \multicolumn{1}{c}{\begin{tabular}[c]{@{}c@{}}Saccade\\ velocity\\ head\end{tabular}} &
  \multicolumn{1}{c}{\begin{tabular}[c]{@{}c@{}}Saccade\\ velocity\\ gaze\end{tabular}} &
  \multicolumn{1}{c}{Coefficient K} \\ \hline
\textit{\textbf{Simple}} &
  \textit{\textbf{Large}} &
  \textit{\textbf{Dense}} &
  \begin{tabular}[c]{@{}l@{}}p < .001\\ d = -0.217\end{tabular} &
  \begin{tabular}[c]{@{}l@{}}p = 0.003\\ d = -0.142\end{tabular} &
  \begin{tabular}[c]{@{}l@{}}p < .001\\ d = 0.536\end{tabular} &
  \begin{tabular}[c]{@{}l@{}}p = 0.006\\ d = -0.13\end{tabular} &
  \begin{tabular}[c]{@{}l@{}}p = 0.025\\ d = -0.103\end{tabular} &
  \begin{tabular}[c]{@{}l@{}}p < .001\\ d = 0.577\end{tabular} &
  \begin{tabular}[c]{@{}l@{}}p = 0.107\\ d = 0.577\end{tabular} &
  \begin{tabular}[c]{@{}l@{}}p < .001\\ d = 0.362\end{tabular} \\
\textit{\textbf{Simple}} &
  \textit{\textbf{Large}} &
  \textit{\textbf{Wide}} &
  \begin{tabular}[c]{@{}l@{}}p = 0.036\\ d = 0.09\end{tabular} &
  \begin{tabular}[c]{@{}l@{}}p < .001\\ d = 0.584\end{tabular} &
  \begin{tabular}[c]{@{}l@{}}p = 0.004\\ d = 0.127\end{tabular} &
  \begin{tabular}[c]{@{}l@{}}p < .001\\ d = 0.62\end{tabular} &
  \begin{tabular}[c]{@{}l@{}}p < .001\\ d = 0.328\end{tabular} &
  \begin{tabular}[c]{@{}l@{}}p < .001\\ d = 0.186\end{tabular} &
  \begin{tabular}[c]{@{}l@{}}p < .001\\ d = 0.513\end{tabular} &
  \begin{tabular}[c]{@{}l@{}}p < .001\\ d = -0.498\end{tabular} \\
\textit{\textbf{Simple}} &
  \textit{\textbf{Small}} &
  \textit{\textbf{Dense}} &
  \begin{tabular}[c]{@{}l@{}}p = 0.476\\ d = 0.034\end{tabular} &
  \begin{tabular}[c]{@{}l@{}}p = 0.198\\ d = -0.06\end{tabular} &
  \begin{tabular}[c]{@{}l@{}}p < .001\\ d = 0.311\end{tabular} &
  \begin{tabular}[c]{@{}l@{}}p = 0.379\\ d = -0.042\end{tabular} &
  \begin{tabular}[c]{@{}l@{}}p = 0.045\\ d = 0.089\end{tabular} &
  \begin{tabular}[c]{@{}l@{}}p < .001\\ d = 0.376\end{tabular} &
  \begin{tabular}[c]{@{}l@{}}p = 0.02\\ d = 0.104\end{tabular} &
  \begin{tabular}[c]{@{}l@{}}p < .001\\ d = 0.346\end{tabular} \\
\textit{\textbf{Simple}} &
  \textit{\textbf{Small}} &
  \textit{\textbf{Wide}} &
  \begin{tabular}[c]{@{}l@{}}p < .001\\ d = 0.518\end{tabular} &
  \begin{tabular}[c]{@{}l@{}}p < .001\\ d = 0.828\end{tabular} &
  \begin{tabular}[c]{@{}l@{}}p < .001\\ d = 0.411\end{tabular} &
  \begin{tabular}[c]{@{}l@{}}p < .001\\ d = 0.871\end{tabular} &
  \begin{tabular}[c]{@{}l@{}}p < .001\\ d = 0.623\end{tabular} &
  \begin{tabular}[c]{@{}l@{}}p < .001\\ d = 0.504\end{tabular} &
  \begin{tabular}[c]{@{}l@{}}p < .001\\ d = 0.885\end{tabular} &
  \begin{tabular}[c]{@{}l@{}}p < .001\\ d = -0.662\end{tabular} \\
\textit{\textbf{Complex}} &
  \textit{\textbf{Large}} &
  \textit{\textbf{Dense}} &
  \begin{tabular}[c]{@{}l@{}}p = 0.241\\ d = 0.048\end{tabular} &
  \begin{tabular}[c]{@{}l@{}}p < .001\\ d = 0.197\end{tabular} &
  \begin{tabular}[c]{@{}l@{}}p < .001\\ d = 0.385\end{tabular} &
  \begin{tabular}[c]{@{}l@{}}p < .001\\ d = 0.198\end{tabular} &
  \begin{tabular}[c]{@{}l@{}}p < .001\\ d = 0.246\end{tabular} &
  \begin{tabular}[c]{@{}l@{}}p < .001\\ d = 0.4\end{tabular} &
  \begin{tabular}[c]{@{}l@{}}p < .001\\ d = 0.262\end{tabular} &
  \begin{tabular}[c]{@{}l@{}}p = 0.038\\ d = -0.083\end{tabular} \\
\textit{\textbf{Complex}} &
  \textit{\textbf{Large}} &
  \textit{\textbf{Wide}} &
  \begin{tabular}[c]{@{}l@{}}p = 0.754\\ d = -0.013\end{tabular} &
  \begin{tabular}[c]{@{}l@{}}p = 0.386\\ d = -0.034\end{tabular} &
  \begin{tabular}[c]{@{}l@{}}p < .001\\ d = 0.827\end{tabular} &
  \begin{tabular}[c]{@{}l@{}}p = 0.338\\ d = 0.038\end{tabular} &
  \begin{tabular}[c]{@{}l@{}}p = 0.004\\ d = -0.117\end{tabular} &
  \begin{tabular}[c]{@{}l@{}}p < .001\\ d = 1.034\end{tabular} &
  \begin{tabular}[c]{@{}l@{}}p = 0.342\\ d = -0.037\end{tabular} &
  \begin{tabular}[c]{@{}l@{}}p = 0.172\\ d = -0.054\end{tabular} \\
\textit{\textbf{Complex}} &
  \textit{\textbf{Small}} &
  \textit{\textbf{Dense}} &
  \begin{tabular}[c]{@{}l@{}}p = 0.004\\ d = -0.107\end{tabular} &
  \begin{tabular}[c]{@{}l@{}}p = 0.008\\ d = 0.098\end{tabular} &
  \begin{tabular}[c]{@{}l@{}}p < .001\\ d = 0.383\end{tabular} &
  \begin{tabular}[c]{@{}l@{}}p = 0.004\\ d = 0.106\end{tabular} &
  \begin{tabular}[c]{@{}l@{}}p = 0.003\\ d = 0.108\end{tabular} &
  \begin{tabular}[c]{@{}l@{}}p < .001\\ d = 0.439\end{tabular} &
  \begin{tabular}[c]{@{}l@{}}p < .001\\ d = 0.132\end{tabular} &
  \begin{tabular}[c]{@{}l@{}}p = 0.008\\ d = 0.091\end{tabular} \\
\textit{\textbf{Complex}} &
  \textit{\textbf{Small}} &
  \textit{\textbf{Wide}} &
  \begin{tabular}[c]{@{}l@{}}p < .001\\ d = 0.206\end{tabular} &
  \begin{tabular}[c]{@{}l@{}}p = 0.79\\ d = -0.011\end{tabular} &
  \begin{tabular}[c]{@{}l@{}}p < .001\\ d = 0.942\end{tabular} &
  \begin{tabular}[c]{@{}l@{}}p < .001\\ d = 0.14\end{tabular} &
  \begin{tabular}[c]{@{}l@{}}p < .001\\ d = -0.168\end{tabular} &
  \begin{tabular}[c]{@{}l@{}}p < .001\\ d = 1.023\end{tabular} &
  \begin{tabular}[c]{@{}l@{}}p = 0.366\\ d = -0.038\end{tabular} &
  \begin{tabular}[c]{@{}l@{}}p = 0.274\\ d = 0.044\end{tabular} \\ \hline
\multicolumn{3}{c}{\textbf{Total Count}} &
  5 &
  5 &
  \textbf{8} &
  \textbf{6} &
  \textbf{8} &
  \textbf{8} &
  5 &
  \textbf{6} \\ \hline
\end{tabular}%
}
\end{table}

The features calculated through the gaze data pipeline are as follows, with references to George and Routray et al.~\cite{george2016score} and Krejtz et al.~\cite{krejtz2016:cok}: Fixation Duration, Fixation std x, Fixation std y, Fixation Velocity, Saccade Duration, Saccade Amplitude Eye, Saccade Amplitude Head, Saccade Amplitude Gaze, Saccade Velocity Eye, Saccade Velocity Head, Saccade Velocity Gaze, and Coefficient K. For Coefficient K, fixed values of fixation duration and saccade amplitude were used during the z-score transformation process.

Next, we proceed the selecting features (observation data) that are suitable for predicting selection intention from gaze-related features. To do this, we classified the collected data into True and False based on selection intention and then selected features that exhibited different characteristics between classes (confirmed through independent t-tests). This process was performed separately for each condition. Before conducting the t-tests, we checked the kurtosis and skewness to verify normality, and if equality of variances was rejected, we applied Welch correction to adjust the degrees of freedom and the p-value. Detailed results are shown in the Table~\ref{tab:metric:selection:part1} and~\ref{tab:metric:selection:part2}. Features where statistically significant differences between class types were found in 6 or more out of 8 conditions were selected as observation data.

\subsection{Fitting}

Next, we proceeded with fitting the selected observation data to probability density functions. This step is necessary to convert new observation data into posterior data for distinguishing selection intention. For the N selected observation data, a total of 2N fittings were performed according to class type, and these were used to construct the Bayesian model. The functions used for fitting include Normal, Gamma, Inverse Gamma, Exponential, Log-normal, Logistic, Log-logistic, Frechet, Gumbel, Gompertz, Weibull, Beta Prime, Cauchy, Chi-squared, and F-distribution. Parameters for each function were estimated using the maximum likelihood estimation method, and a function was considered properly fitted if the result of the Kolmogorov-Smirnov test showed a p-value greater than 0.05. In cases where more than one function had a p-value greater than 0.05, the function with the higher p-value was chosen. If the data sample size exceeded 5,000, the test could not be conducted, and the data was assumed to follow a normal distribution for fitting purposes. In cases where fitting was not possible, the feature was labeled as "None" and a uniform distribution was applied. Features that could not be fitted in any condition were not included in the analysis results presented in this section. The fitting results can be found in the Table~\ref{tab:fitting:fixationvelocity} to~\ref{tab:fitting:coefficientk}.

\newpage
\begin{table}[ht]
\caption{The fitting for the 1st Fixation Velocity feature was conducted independently for each condition and selection class. The results of the fitting are displayed for each condition, with fitting performed across multiple distribution functions such as Normal, Gamma, Inverse Gamma, and others. If fitting was unsuccessful, the feature was labeled as "None" and modeled using a uniform distribution.}
\label{tab:fitting:fixationvelocity}
\begin{tabular}{cccc|clll}
\hline
\textbf{\begin{tabular}[c]{@{}c@{}}Task \\ Type\end{tabular}} &
  \textbf{\begin{tabular}[c]{@{}c@{}}Target \\ Width\end{tabular}} &
  \textbf{\begin{tabular}[c]{@{}c@{}}Target \\ Density\end{tabular}} &
  \textbf{Selection} &
  \textbf{Function} &
  \multicolumn{1}{c}{\textbf{Parameter 1}} &
  \multicolumn{1}{c}{\textbf{Parameter 2}} &
  \multicolumn{1}{c}{\textbf{p-value}} \\ \hline
\multirow{8}{*}{\textit{\textbf{Simple}}} &
  \multirow{4}{*}{\textit{\textbf{Large}}} &
  \multirow{2}{*}{\textit{\textbf{Dense}}} &
  \textit{\textbf{True}} &
  Weibull &
  1.397 &
  0.014 &
  0.413 \\ \cline{4-8} 
 &
   &
   &
  \textit{\textbf{False}} &
  Gamma &
  2.311 &
  0.006 &
  0.423 \\ \cline{3-8} 
 &
   &
  \multirow{2}{*}{\textit{\textbf{Wide}}} &
  \textit{\textbf{True}} &
  Weibull &
  2.22 &
  0.034 &
  0.966 \\ \cline{4-8} 
 &
   &
   &
  \textit{\textbf{False}} &
  Weibull &
  1.844 &
  0.03 &
  0.61 \\ \cline{2-8} 
 &
  \multirow{4}{*}{\textit{\textbf{Small}}} &
  \multirow{2}{*}{\textit{\textbf{Dense}}} &
  \textit{\textbf{True}} &
  Weibull &
  1.492 &
  0.015 &
  0.844 \\ \cline{4-8} 
 &
   &
   &
  \textit{\textbf{False}} &
  Log-normal &
  -4.527 &
  0.753 &
  0.125 \\ \cline{3-8} 
 &
   &
  \multirow{2}{*}{\textit{\textbf{Wide}}} &
  \textit{\textbf{True}} &
  Normal &
  0.032 &
  2.096×10-4 &
  0.858 \\ \cline{4-8} 
 &
   &
   &
  \textit{\textbf{False}} &
  Weibull &
  1.727 &
  0.029 &
  0.887 \\ \hline
\multirow{8}{*}{\textit{\textbf{Complex}}} &
  \multirow{4}{*}{\textit{\textbf{Large}}} &
  \multirow{2}{*}{\textit{\textbf{Dense}}} &
  \textit{\textbf{True}} &
  Gamma &
  2.661 &
  0.004 &
  0.697 \\ \cline{4-8} 
 &
   &
   &
  \textit{\textbf{False}} &
  Normal &
  0.013 &
  4.576×10-5 &
  - \\ \cline{3-8} 
 &
   &
  \multirow{2}{*}{\textit{\textbf{Wide}}} &
  \textit{\textbf{True}} &
  Weibull &
  2.191 &
  0.02 &
  0.965 \\ \cline{4-8} 
 &
   &
   &
  \textit{\textbf{False}} &
  Normal &
  0.022 &
  1.024×10-4 &
  - \\ \cline{2-8} 
 &
  \multirow{4}{*}{\textit{\textbf{Small}}} &
  \multirow{2}{*}{\textit{\textbf{Dense}}} &
  \textit{\textbf{True}} &
  Gamma &
  2.825 &
  0.004 &
  0.376 \\ \cline{4-8} 
 &
   &
   &
  \textit{\textbf{False}} &
  Normal &
  0.013 &
  4.617×10-5 &
  - \\ \cline{3-8} 
 &
   &
  \multirow{2}{*}{\textit{\textbf{Wide}}} &
  \textit{\textbf{True}} &
  Weibull &
  1.838 &
  0.022 &
  0.819 \\ \cline{4-8} 
 &
   &
   &
  \textit{\textbf{False}} &
  Normal &
  0.023 &
  1.092×10-4 &
  - \\ \hline
\end{tabular}
\end{table}

\begin{table}[ht]
\caption{The fitting for the 1st Fixation std x feature was conducted independently for each condition and selection class. The results of the fitting are displayed for each condition, with fitting performed across multiple distribution functions such as Normal, Gamma, Inverse Gamma, and others. If fitting was unsuccessful, the feature was labeled as "None" and modeled using a uniform distribution.}
\label{tab:fitting:fixationstdx}
\begin{tabular}{cccc|llll}
\hline
\textbf{\begin{tabular}[c]{@{}c@{}}Task \\ Type\end{tabular}} &
  \textbf{\begin{tabular}[c]{@{}c@{}}Target \\ Width\end{tabular}} &
  \textbf{\begin{tabular}[c]{@{}c@{}}Target \\ Density\end{tabular}} &
  \textbf{Selection} &
  \multicolumn{1}{c}{\textbf{Function}} &
  \multicolumn{1}{c}{\textbf{Parameter 1}} &
  \multicolumn{1}{c}{\textbf{Parameter 2}} &
  \multicolumn{1}{c}{\textbf{p-value}} \\ \hline
\multirow{8}{*}{\textit{\textbf{Simple}}} &
  \multirow{4}{*}{\textit{\textbf{Large}}} &
  \multirow{2}{*}{\textit{\textbf{Dense}}} &
  \textit{\textbf{True}} &
  Gompertz &
  0.958 &
  2.647 &
  0.545 \\ \cline{4-8} 
 &
   &
   &
  \textit{\textbf{False}} &
  Weibull &
  1.441 &
  0.324 &
  0.188 \\ \cline{3-8} 
 &
   &
  \multirow{2}{*}{\textit{\textbf{Wide}}} &
  \textit{\textbf{True}} &
  Weibull &
  1.675 &
  0.482 &
  0.685 \\ \cline{4-8} 
 &
   &
   &
  \textit{\textbf{False}} &
  Weibull &
  1.508 &
  0.436 &
  0.409 \\ \cline{2-8} 
 &
  \multirow{4}{*}{\textit{\textbf{Small}}} &
  \multirow{2}{*}{\textit{\textbf{Dense}}} &
  \textit{\textbf{True}} &
  Gompertz &
  1.054 &
  2.464 &
  0.343 \\ \cline{4-8} 
 &
   &
   &
  \textit{\textbf{False}} &
  Weibull &
  1.522 &
  0.308 &
  0.409 \\ \cline{3-8} 
 &
   &
  \multirow{2}{*}{\textit{\textbf{Wide}}} &
  \textit{\textbf{True}} &
  Gompertz &
  0.49 &
  2.094 &
  0.658 \\ \cline{4-8} 
 &
   &
   &
  \textit{\textbf{False}} &
  Gompertz &
  0.658 &
  2.221 &
  0.285 \\ \hline
\multirow{8}{*}{\textit{\textbf{Complex}}} &
  \multirow{4}{*}{\textit{\textbf{Large}}} &
  \multirow{2}{*}{\textit{\textbf{Dense}}} &
  \textit{\textbf{True}} &
  Weibull &
  1.432 &
  0.3 &
  0.778 \\ \cline{4-8} 
 &
   &
   &
  \textit{\textbf{False}} &
  Normal &
  0.294 &
  0.035 &
  - \\ \cline{3-8} 
 &
   &
  \multirow{2}{*}{\textit{\textbf{Wide}}} &
  \textit{\textbf{True}} &
  Weibull &
  1.543 &
  0.426 &
  0.738 \\ \cline{4-8} 
 &
   &
   &
  \textit{\textbf{False}} &
  Normal &
  0.411 &
  0.055 &
  - \\ \cline{2-8} 
 &
  \multirow{4}{*}{\textit{\textbf{Small}}} &
  \multirow{2}{*}{\textit{\textbf{Dense}}} &
  \textit{\textbf{True}} &
  Weibull &
  1.487 &
  0.308 &
  0.708 \\ \cline{4-8} 
 &
   &
   &
  \textit{\textbf{False}} &
  Normal &
  0.3 &
  0.035 &
  - \\ \cline{3-8} 
 &
   &
  \multirow{2}{*}{\textit{\textbf{Wide}}} &
  \textit{\textbf{True}} &
  Weibull &
  1.396 &
  0.422 &
  0.588 \\ \cline{4-8} 
 &
   &
   &
  \textit{\textbf{False}} &
  Normal &
  0.421 &
  0.054 &
  - \\ \hline
\end{tabular}
\end{table}

\newpage

\begin{table}[ht]
\caption{The fitting for the 1st Fixation std y feature was conducted independently for each condition and selection class. The results of the fitting are displayed for each condition, with fitting performed across multiple distribution functions such as Normal, Gamma, Inverse Gamma, and others. If fitting was unsuccessful, the feature was labeled as "None" and modeled using a uniform distribution.}
\label{tab:fitting:fixationstdy}
\begin{tabular}{cccc|llll}
\hline
\textbf{\begin{tabular}[c]{@{}c@{}}Task \\ Type\end{tabular}} &
  \textbf{\begin{tabular}[c]{@{}c@{}}Target \\ Width\end{tabular}} &
  \textbf{\begin{tabular}[c]{@{}c@{}}Target \\ Density\end{tabular}} &
  \textbf{Selection} &
  \multicolumn{1}{c}{\textbf{Function}} &
  \multicolumn{1}{c}{\textbf{Parameter 1}} &
  \multicolumn{1}{c}{\textbf{Parameter 2}} &
  \multicolumn{1}{c}{\textbf{p-value}} \\ \hline
\multirow{8}{*}{\textit{\textbf{Simple}}} &
  \multirow{4}{*}{\textit{\textbf{Large}}} &
  \multirow{2}{*}{\textit{\textbf{Dense}}} &
  \textit{\textbf{True}} &
  Normal &
  0.194 &
  0.016 &
  0.07 \\ \cline{4-8} 
 &
   &
   &
  \textit{\textbf{False}} &
  Weibull &
  1.547 &
  0.271 &
  0.464 \\ \cline{3-8} 
 &
   &
  \multirow{2}{*}{\textit{\textbf{Wide}}} &
  \textit{\textbf{True}} &
  Weibull &
  1.656 &
  0.345 &
  0.889 \\ \cline{4-8} 
 &
   &
   &
  \textit{\textbf{False}} &
  Gompertz &
  0.628 &
  2.822 &
  0.368 \\ \cline{2-8} 
 &
  \multirow{4}{*}{\textit{\textbf{Small}}} &
  \multirow{2}{*}{\textit{\textbf{Dense}}} &
  \textit{\textbf{True}} &
  Gompertz &
  0.96 &
  2.822 &
  0.839 \\ \cline{4-8} 
 &
   &
   &
  \textit{\textbf{False}} &
  Weibull &
  1.529 &
  0.26 &
  0.659 \\ \cline{3-8} 
 &
   &
  \multirow{2}{*}{\textit{\textbf{Wide}}} &
  \textit{\textbf{True}} &
  Gompertz &
  0.623 &
  2.497 &
  0.912 \\ \cline{4-8} 
 &
   &
   &
  \textit{\textbf{False}} &
  Gompertz &
  0.551 &
  2.903 &
  0.355 \\ \hline
\multirow{8}{*}{\textit{\textbf{Complex}}} &
  \multirow{4}{*}{\textit{\textbf{Large}}} &
  \multirow{2}{*}{\textit{\textbf{Dense}}} &
  \textit{\textbf{True}} &
  Weibull &
  1.45 &
  0.232 &
  0.578 \\ \cline{4-8} 
 &
   &
   &
  \textit{\textbf{False}} &
  Normal &
  0.229 &
  0.022 &
  - \\ \cline{3-8} 
 &
   &
  \multirow{2}{*}{\textit{\textbf{Wide}}} &
  \textit{\textbf{True}} &
  Weibull &
  1.625 &
  0.298 &
  0.56 \\ \cline{4-8} 
 &
   &
   &
  \textit{\textbf{False}} &
  Normal &
  0.295 &
  0.031 &
  - \\ \cline{2-8} 
 &
  \multirow{4}{*}{\textit{\textbf{Small}}} &
  \multirow{2}{*}{\textit{\textbf{Dense}}} &
  \textit{\textbf{True}} &
  Weibull &
  1.5 &
  0.243 &
  0.455 \\ \cline{4-8} 
 &
   &
   &
  \textit{\textbf{False}} &
  Normal &
  0.244 &
  0.024 &
  - \\ \cline{3-8} 
 &
   &
  \multirow{2}{*}{\textit{\textbf{Wide}}} &
  \textit{\textbf{True}} &
  Weibull &
  1.302 &
  0.261 &
  0.839 \\ \cline{4-8} 
 &
   &
   &
  \textit{\textbf{False}} &
  Normal &
  0.286 &
  0.03 &
  - \\ \hline
\end{tabular}
\end{table}

\begin{table}[ht]
\caption{The fitting for the 2nd Fixation Velocity feature was conducted independently for each condition and selection class. The results of the fitting are displayed for each condition, with fitting performed across multiple distribution functions such as Normal, Gamma, Inverse Gamma, and others. If fitting was unsuccessful, the feature was labeled as "None" and modeled using a uniform distribution.}
\label{tab:fitting:currentfixationvelocity}
\begin{tabular}{cccc|llll}
\hline
\textbf{\begin{tabular}[c]{@{}c@{}}Task \\ Type\end{tabular}} &
  \textbf{\begin{tabular}[c]{@{}c@{}}Target \\ Width\end{tabular}} &
  \textbf{\begin{tabular}[c]{@{}c@{}}Target \\ Density\end{tabular}} &
  \textbf{Selection} &
  \multicolumn{1}{c}{\textbf{Function}} &
  \multicolumn{1}{c}{\textbf{Parameter 1}} &
  \multicolumn{1}{c}{\textbf{Parameter 2}} &
  \multicolumn{1}{c}{\textbf{p-value}} \\ \hline
\multirow{8}{*}{\textit{\textbf{Simple}}} &
  \multirow{4}{*}{\textit{\textbf{Large}}} &
  \multirow{2}{*}{\textit{\textbf{Dense}}} &
  \textit{\textbf{True}} &
  Log-normal &
  -4.636 &
  0.711 &
  0.466 \\ \cline{4-8} 
 &
   &
   &
  \textit{\textbf{False}} &
  Normal &
  0.019 &
  1.491×10-4 &
  0.141 \\ \cline{3-8} 
 &
   &
  \multirow{2}{*}{\textit{\textbf{Wide}}} &
  \textit{\textbf{True}} &
  Gamma &
  3.652 &
  0.007 &
  0.722 \\ \cline{4-8} 
 &
   &
   &
  \textit{\textbf{False}} &
  Weibull &
  1.933 &
  0.033 &
  0.453 \\ \cline{2-8} 
 &
  \multirow{4}{*}{\textit{\textbf{Small}}} &
  \multirow{2}{*}{\textit{\textbf{Dense}}} &
  \textit{\textbf{True}} &
  Log-normal &
  -4.472 &
  0.576 &
  0.128 \\ \cline{4-8} 
 &
   &
   &
  \textit{\textbf{False}} &
  Logistic &
  0.019 &
  0.008 &
  0.089 \\ \cline{3-8} 
 &
   &
  \multirow{2}{*}{\textit{\textbf{Wide}}} &
  \textit{\textbf{True}} &
  Weibull &
  2.304 &
  0.026 &
  0.827 \\ \cline{4-8} 
 &
   &
   &
  \textit{\textbf{False}} &
  Weibull &
  1.938 &
  0.033 &
  0.125 \\ \hline
\multirow{8}{*}{\textit{\textbf{Complex}}} &
  \multirow{4}{*}{\textit{\textbf{Large}}} &
  \multirow{2}{*}{\textit{\textbf{Dense}}} &
  \textit{\textbf{True}} &
  Gamma &
  3.182 &
  0.004 &
  0.814 \\ \cline{4-8} 
 &
   &
   &
  \textit{\textbf{False}} &
  Normal &
  0.011 &
  5.443×10-5 &
  - \\ \cline{3-8} 
 &
   &
  \multirow{2}{*}{\textit{\textbf{Wide}}} &
  \textit{\textbf{True}} &
  Gamma &
  4.07 &
  0.004 &
  0.809 \\ \cline{4-8} 
 &
   &
   &
  \textit{\textbf{False}} &
  Normal &
  0.021 &
  1.267×10-4 &
  - \\ \cline{2-8} 
 &
  \multirow{4}{*}{\textit{\textbf{Small}}} &
  \multirow{2}{*}{\textit{\textbf{Dense}}} &
  \textit{\textbf{True}} &
  Log-normal &
  -4.622 &
  0.541 &
  0.415 \\ \cline{4-8} 
 &
   &
   &
  \textit{\textbf{False}} &
  Normal &
  0.011 &
  5.546×10-5 &
  - \\ \cline{3-8} 
 &
   &
  \multirow{2}{*}{\textit{\textbf{Wide}}} &
  \textit{\textbf{True}} &
  Gamma &
  3.999 &
  0.004 &
  0.41 \\ \cline{4-8} 
 &
   &
   &
  \textit{\textbf{False}} &
  Normal &
  0.022 &
  1.350×10-4 &
  - \\ \hline
\end{tabular}
\end{table}

\newpage

\begin{table}[ht]
\caption{The fitting for the 2nd Fixation Duration feature was conducted independently for each condition and selection class. The results of the fitting are displayed for each condition, with fitting performed across multiple distribution functions such as Normal, Gamma, Inverse Gamma, and others. If fitting was unsuccessful, the feature was labeled as "None" and modeled using a uniform distribution.}
\label{tab:fitting:currentfixationduration}
\begin{tabular}{cccc|llll}
\hline
\textbf{\begin{tabular}[c]{@{}c@{}}Task \\ Type\end{tabular}} &
  \textbf{\begin{tabular}[c]{@{}c@{}}Target \\ Width\end{tabular}} &
  \textbf{\begin{tabular}[c]{@{}c@{}}Target \\ Density\end{tabular}} &
  \textbf{Selection} &
  \multicolumn{1}{c}{\textbf{Function}} &
  \multicolumn{1}{c}{\textbf{Parameter 1}} &
  \multicolumn{1}{c}{\textbf{Parameter 2}} &
  \multicolumn{1}{c}{\textbf{p-value}} \\ \hline
\multirow{8}{*}{\textit{\textbf{Simple}}} &
  \multirow{4}{*}{\textit{\textbf{Large}}} &
  \multirow{2}{*}{\textit{\textbf{Dense}}} &
  \textit{\textbf{True}} &
  None &
  0 &
  0 &
  - \\ \cline{4-8} 
 &
   &
   &
  \textit{\textbf{False}} &
  None &
  0 &
  0 &
  - \\ \cline{3-8} 
 &
   &
  \multirow{2}{*}{\textit{\textbf{Wide}}} &
  \textit{\textbf{True}} &
  None &
  0 &
  0 &
  - \\ \cline{4-8} 
 &
   &
   &
  \textit{\textbf{False}} &
  None &
  0 &
  0 &
  - \\ \cline{2-8} 
 &
  \multirow{4}{*}{\textit{\textbf{Small}}} &
  \multirow{2}{*}{\textit{\textbf{Dense}}} &
  \textit{\textbf{True}} &
  None &
  0 &
  0 &
  - \\ \cline{4-8} 
 &
   &
   &
  \textit{\textbf{False}} &
  None &
  0 &
  0 &
  - \\ \cline{3-8} 
 &
   &
  \multirow{2}{*}{\textit{\textbf{Wide}}} &
  \textit{\textbf{True}} &
  None &
  0 &
  0 &
  - \\ \cline{4-8} 
 &
   &
   &
  \textit{\textbf{False}} &
  None &
  0 &
  0 &
  - \\ \hline
\multirow{8}{*}{\textit{\textbf{Complex}}} &
  \multirow{4}{*}{\textit{\textbf{Large}}} &
  \multirow{2}{*}{\textit{\textbf{Dense}}} &
  \textit{\textbf{True}} &
  Gamma &
  1.541 &
  164.248 &
  0.271 \\ \cline{4-8} 
 &
   &
   &
  \textit{\textbf{False}} &
  Normal &
  184.752 &
  8117.234 &
  - \\ \cline{3-8} 
 &
   &
  \multirow{2}{*}{\textit{\textbf{Wide}}} &
  \textit{\textbf{True}} &
  Weibull &
  1.131 &
  214.845 &
  0.266 \\ \cline{4-8} 
 &
   &
   &
  \textit{\textbf{False}} &
  Normal &
  147.648 &
  7237.248 &
  - \\ \cline{2-8} 
 &
  \multirow{4}{*}{\textit{\textbf{Small}}} &
  \multirow{2}{*}{\textit{\textbf{Dense}}} &
  \textit{\textbf{True}} &
  Weibull &
  1.373 &
  352.295 &
  0.57 \\ \cline{4-8} 
 &
   &
   &
  \textit{\textbf{False}} &
  Normal &
  207.833 &
  11236.47 &
  - \\ \cline{3-8} 
 &
   &
  \multirow{2}{*}{\textit{\textbf{Wide}}} &
  \textit{\textbf{True}} &
  Gamma &
  1.342 &
  182.328 &
  0.124 \\ \cline{4-8} 
 &
   &
   &
  \textit{\textbf{False}} &
  Normal &
  152.983 &
  8792.361 &
  - \\ \hline
\end{tabular}
\end{table}

\begin{table}[ht]
\caption{The fitting for the 2nd Fixation std y feature was conducted independently for each condition and selection class. The results of the fitting are displayed for each condition, with fitting performed across multiple distribution functions such as Normal, Gamma, Inverse Gamma, and others. If fitting was unsuccessful, the feature was labeled as "None" and modeled using a uniform distribution.}
\label{tab:fitting:currentfixationstdy}
\begin{tabular}{cccc|llll}
\hline
\textbf{\begin{tabular}[c]{@{}c@{}}Task \\ Type\end{tabular}} &
  \textbf{\begin{tabular}[c]{@{}c@{}}Target \\ Width\end{tabular}} &
  \textbf{\begin{tabular}[c]{@{}c@{}}Target \\ Density\end{tabular}} &
  \textbf{Selection} &
  \multicolumn{1}{c}{\textbf{Function}} &
  \multicolumn{1}{c}{\textbf{Parameter 1}} &
  \multicolumn{1}{c}{\textbf{Parameter 2}} &
  \multicolumn{1}{c}{\textbf{p-value}} \\ \hline
\multirow{8}{*}{\textit{\textbf{Simple}}} &
  \multirow{4}{*}{\textit{\textbf{Large}}} &
  \multirow{2}{*}{\textit{\textbf{Dense}}} &
  \textit{\textbf{True}} &
  Gamma &
  2.116 &
  0.115 &
  0.78 \\ \cline{4-8} 
 &
   &
   &
  \textit{\textbf{False}} &
  Weibull &
  1.178 &
  0.218 &
  0.032 \\ \cline{3-8} 
 &
   &
  \multirow{2}{*}{\textit{\textbf{Wide}}} &
  \textit{\textbf{True}} &
  Weibull &
  1.668 &
  0.455 &
  0.374 \\ \cline{4-8} 
 &
   &
   &
  \textit{\textbf{False}} &
  Gompertz &
  0.703 &
  2.703 &
  0.391 \\ \cline{2-8} 
 &
  \multirow{4}{*}{\textit{\textbf{Small}}} &
  \multirow{2}{*}{\textit{\textbf{Dense}}} &
  \textit{\textbf{True}} &
  Gamma &
  2.467 &
  0.125 &
  0.782 \\ \cline{4-8} 
 &
   &
   &
  \textit{\textbf{False}} &
  Weibull &
  1.103 &
  0.199 &
  0.035 \\ \cline{3-8} 
 &
   &
  \multirow{2}{*}{\textit{\textbf{Wide}}} &
  \textit{\textbf{True}} &
  Weibull &
  1.606 &
  0.483 &
  0.932 \\ \cline{4-8} 
 &
   &
   &
  \textit{\textbf{False}} &
  Gompertz &
  0.608 &
  2.775 &
  0.734 \\ \hline
\multirow{8}{*}{\textit{\textbf{Complex}}} &
  \multirow{4}{*}{\textit{\textbf{Large}}} &
  \multirow{2}{*}{\textit{\textbf{Dense}}} &
  \textit{\textbf{True}} &
  Gamma &
  2.049 &
  0.128 &
  0.587 \\ \cline{4-8} 
 &
   &
   &
  \textit{\textbf{False}} &
  Normal &
  0.227 &
  0.021 &
  - \\ \cline{3-8} 
 &
   &
  \multirow{2}{*}{\textit{\textbf{Wide}}} &
  \textit{\textbf{True}} &
  Weibull &
  1.605 &
  0.333 &
  0.888 \\ \cline{4-8} 
 &
   &
   &
  \textit{\textbf{False}} &
  Normal &
  0.294 &
  0.031 &
  - \\ \cline{2-8} 
 &
  \multirow{4}{*}{\textit{\textbf{Small}}} &
  \multirow{2}{*}{\textit{\textbf{Dense}}} &
  \textit{\textbf{True}} &
  Gamma &
  2.046 &
  0.144 &
  0.658 \\ \cline{4-8} 
 &
   &
   &
  \textit{\textbf{False}} &
  Normal &
  0.242 &
  0.024 &
  - \\ \cline{3-8} 
 &
   &
  \multirow{2}{*}{\textit{\textbf{Wide}}} &
  \textit{\textbf{True}} &
  Weibull &
  1.715 &
  0.357 &
  0.949 \\ \cline{4-8} 
 &
   &
   &
  \textit{\textbf{False}} &
  Normal &
  0.285 &
  0.03 &
  - \\ \hline
\end{tabular}
\end{table}

\newpage

\begin{table}[ht]
\caption{The fitting for the Saccade Amplitude Head feature was conducted independently for each condition and selection class. The results of the fitting are displayed for each condition, with fitting performed across multiple distribution functions such as Normal, Gamma, Inverse Gamma, and others. If fitting was unsuccessful, the feature was labeled as "None" and modeled using a uniform distribution.}
\label{tab:fitting:saccadeamplitudehead}
\begin{tabular}{cccc|llll}
\hline
\textbf{\begin{tabular}[c]{@{}c@{}}Task \\ Type\end{tabular}} &
  \textbf{\begin{tabular}[c]{@{}c@{}}Target \\ Width\end{tabular}} &
  \textbf{\begin{tabular}[c]{@{}c@{}}Target \\ Density\end{tabular}} &
  \textbf{Selection} &
  \multicolumn{1}{c}{\textbf{Function}} &
  \multicolumn{1}{c}{\textbf{Parameter 1}} &
  \multicolumn{1}{c}{\textbf{Parameter 2}} &
  \multicolumn{1}{c}{\textbf{p-value}} \\ \hline
\multirow{8}{*}{\textit{\textbf{Simple}}} &
  \multirow{4}{*}{\textit{\textbf{Large}}} &
  \multirow{2}{*}{\textit{\textbf{Dense}}} &
  \textit{\textbf{True}} &
  Log-normal &
  -2.692 &
  1.169 &
  0.297 \\ \cline{4-8} 
 &
   &
   &
  \textit{\textbf{False}} &
  Log-normal &
  -2.319 &
  1.371 &
  0.044 \\ \cline{3-8} 
 &
   &
  \multirow{2}{*}{\textit{\textbf{Wide}}} &
  \textit{\textbf{True}} &
  BetaPrime &
  1.186 &
  3.018 &
  0.091 \\ \cline{4-8} 
 &
   &
   &
  \textit{\textbf{False}} &
  Weibull &
  0.883 &
  0.611 &
  0.032 \\ \cline{2-8} 
 &
  \multirow{4}{*}{\textit{\textbf{Small}}} &
  \multirow{2}{*}{\textit{\textbf{Dense}}} &
  \textit{\textbf{True}} &
  Log-normal &
  -2.145 &
  1.206 &
  0.547 \\ \cline{4-8} 
 &
   &
   &
  \textit{\textbf{False}} &
  Log-normal &
  -2.085 &
  1.32 &
  0.066 \\ \cline{3-8} 
 &
   &
  \multirow{2}{*}{\textit{\textbf{Wide}}} &
  \textit{\textbf{True}} &
  Log-normal &
  -1.428 &
  1.12 &
  0.392 \\ \cline{4-8} 
 &
   &
   &
  \textit{\textbf{False}} &
  Weibull &
  0.852 &
  0.658 &
  0.02 \\ \hline
\multirow{8}{*}{\textit{\textbf{Complex}}} &
  \multirow{4}{*}{\textit{\textbf{Large}}} &
  \multirow{2}{*}{\textit{\textbf{Dense}}} &
  \textit{\textbf{True}} &
  Log-normal &
  -2.752 &
  1.111 &
  0.641 \\ \cline{4-8} 
 &
   &
   &
  \textit{\textbf{False}} &
  Normal &
  0.164 &
  0.024 &
  - \\ \cline{3-8} 
 &
   &
  \multirow{2}{*}{\textit{\textbf{Wide}}} &
  \textit{\textbf{True}} &
  Log-normal &
  -1.814 &
  1.054 &
  0.485 \\ \cline{4-8} 
 &
   &
   &
  \textit{\textbf{False}} &
  Normal &
  0.557 &
  0.231 &
  - \\ \cline{2-8} 
 &
  \multirow{4}{*}{\textit{\textbf{Small}}} &
  \multirow{2}{*}{\textit{\textbf{Dense}}} &
  \textit{\textbf{True}} &
  Log-normal &
  -2.726 &
  1.007 &
  0.426 \\ \cline{4-8} 
 &
   &
   &
  \textit{\textbf{False}} &
  Normal &
  0.157 &
  0.022 &
  - \\ \cline{3-8} 
 &
   &
  \multirow{2}{*}{\textit{\textbf{Wide}}} &
  \textit{\textbf{True}} &
  Gamma &
  1.289 &
  0.161 &
  0.994 \\ \cline{4-8} 
 &
   &
   &
  \textit{\textbf{False}} &
  Normal &
  0.509 &
  0.194 &
  - \\ \hline
\end{tabular}
\end{table}

\begin{table}[ht]
\caption{The fitting for the Saccade Velocity Head feature was conducted independently for each condition and selection class. The results of the fitting are displayed for each condition, with fitting performed across multiple distribution functions such as Normal, Gamma, Inverse Gamma, and others. If fitting was unsuccessful, the feature was labeled as "None" and modeled using a uniform distribution.}
\label{tab:fitting:saccadevelocityhead}
\begin{tabular}{cccc|llll}
\hline
\textbf{\begin{tabular}[c]{@{}c@{}}Task \\ Type\end{tabular}} &
  \textbf{\begin{tabular}[c]{@{}c@{}}Target \\ Width\end{tabular}} &
  \textbf{\begin{tabular}[c]{@{}c@{}}Target \\ Density\end{tabular}} &
  \textbf{Selection} &
  \multicolumn{1}{c}{\textbf{Function}} &
  \multicolumn{1}{c}{\textbf{Parameter 1}} &
  \multicolumn{1}{c}{\textbf{Parameter 2}} &
  \multicolumn{1}{c}{\textbf{p-value}} \\ \hline
\multirow{8}{*}{\textit{\textbf{Simple}}} &
  \multirow{4}{*}{\textit{\textbf{Large}}} &
  \multirow{2}{*}{\textit{\textbf{Dense}}} &
  \textit{\textbf{True}} &
  None &
  0 &
  0 &
  - \\ \cline{4-8} 
 &
   &
   &
  \textit{\textbf{False}} &
  None &
  0 &
  0 &
  - \\ \cline{3-8} 
 &
   &
  \multirow{2}{*}{\textit{\textbf{Wide}}} &
  \textit{\textbf{True}} &
  Gamma &
  1.247 &
  0.02 &
  0.481 \\ \cline{4-8} 
 &
   &
   &
  \textit{\textbf{False}} &
  Gompertz &
  2.672 &
  9.941 &
  0.026 \\ \cline{2-8} 
 &
  \multirow{4}{*}{\textit{\textbf{Small}}} &
  \multirow{2}{*}{\textit{\textbf{Dense}}} &
  \textit{\textbf{True}} &
  Exponential &
  94.545 &
  0 &
  0.51 \\ \cline{4-8} 
 &
   &
   &
  \textit{\textbf{False}} &
  Weibull &
  0.926 &
  0.013 &
  0.035 \\ \cline{3-8} 
 &
   &
  \multirow{2}{*}{\textit{\textbf{Wide}}} &
  \textit{\textbf{True}} &
  Gamma &
  1.421 &
  0.014 &
  0.411 \\ \cline{4-8} 
 &
   &
   &
  \textit{\textbf{False}} &
  Gompertz &
  2.471 &
  9.77 &
  0.003 \\ \hline
\multirow{8}{*}{\textit{\textbf{Complex}}} &
  \multirow{4}{*}{\textit{\textbf{Large}}} &
  \multirow{2}{*}{\textit{\textbf{Dense}}} &
  \textit{\textbf{True}} &
  Log-normal &
  -5.792 &
  1.104 &
  0.406 \\ \cline{4-8} 
 &
   &
   &
  \textit{\textbf{False}} &
  Normal &
  0.008 &
  0.00005006 &
  - \\ \cline{3-8} 
 &
   &
  \multirow{2}{*}{\textit{\textbf{Wide}}} &
  \textit{\textbf{True}} &
  Log-normal &
  -4.956 &
  0.933 &
  0.541 \\ \cline{4-8} 
 &
   &
   &
  \textit{\textbf{False}} &
  Normal &
  0.024 &
  0.0002715 &
  - \\ \cline{2-8} 
 &
  \multirow{4}{*}{\textit{\textbf{Small}}} &
  \multirow{2}{*}{\textit{\textbf{Dense}}} &
  \textit{\textbf{True}} &
  Log-normal &
  -5.737 &
  0.983 &
  0.557 \\ \cline{4-8} 
 &
   &
   &
  \textit{\textbf{False}} &
  Normal &
  0.008 &
  0.0000516 &
  - \\ \cline{3-8} 
 &
   &
  \multirow{2}{*}{\textit{\textbf{Wide}}} &
  \textit{\textbf{True}} &
  Weibull &
  1.287 &
  0.012 &
  0.941 \\ \cline{4-8} 
 &
   &
   &
  \textit{\textbf{False}} &
  Normal &
  0.024 &
  0.0002649 &
  - \\ \hline
\end{tabular}
\end{table}

\newpage

\begin{table}[ht]
\caption{The fitting for the Coefficient K feature was conducted independently for each condition and selection class. The results of the fitting are displayed for each condition, with fitting performed across multiple distribution functions such as Normal, Gamma, Inverse Gamma, and others. If fitting was unsuccessful, the feature was labeled as "None" and modeled using a uniform distribution.}
\label{tab:fitting:coefficientk}
\begin{tabular}{cccc|llll}
\hline
\textbf{\begin{tabular}[c]{@{}c@{}}Task \\ Type\end{tabular}} &
  \textbf{\begin{tabular}[c]{@{}c@{}}Target \\ Width\end{tabular}} &
  \textbf{\begin{tabular}[c]{@{}c@{}}Target \\ Density\end{tabular}} &
  \textbf{Selection} &
  \multicolumn{1}{c}{\textbf{Function}} &
  \multicolumn{1}{c}{\textbf{Parameter 1}} &
  \multicolumn{1}{c}{\textbf{Parameter 2}} &
  \multicolumn{1}{c}{\textbf{p-value}} \\ \hline
\multirow{8}{*}{\textit{\textbf{Simple}}} &
  \multirow{4}{*}{\textit{\textbf{Large}}} &
  \multirow{2}{*}{\textit{\textbf{Dense}}} &
  \textit{\textbf{True}} &
  Logistic &
  0.221 &
  0.657 &
  0.325 \\ \cline{4-8} 
 &
   &
   &
  \textit{\textbf{False}} &
  Logistic &
  0.571 &
  0.619 &
  0.185 \\ \cline{3-8} 
 &
   &
  \multirow{2}{*}{\textit{\textbf{Wide}}} &
  \textit{\textbf{True}} &
  Gumbel &
  0.66 &
  0.895 &
  0.166 \\ \cline{4-8} 
 &
   &
   &
  \textit{\textbf{False}} &
  Cauchy &
  0.625 &
  0.57 &
  0.004 \\ \cline{2-8} 
 &
  \multirow{4}{*}{\textit{\textbf{Small}}} &
  \multirow{2}{*}{\textit{\textbf{Dense}}} &
  \textit{\textbf{True}} &
  Logistic &
  0.322 &
  0.637 &
  0.123 \\ \cline{4-8} 
 &
   &
   &
  \textit{\textbf{False}} &
  Logistic &
  0.627 &
  0.645 &
  0.171 \\ \cline{3-8} 
 &
   &
  \multirow{2}{*}{\textit{\textbf{Wide}}} &
  \textit{\textbf{True}} &
  Logistic &
  0.992 &
  0.448 &
  0.012 \\ \cline{4-8} 
 &
   &
   &
  \textit{\textbf{False}} &
  Cauchy &
  0.6 &
  0.621 &
  0.006 \\ \hline
\multirow{8}{*}{\textit{\textbf{Complex}}} &
  \multirow{4}{*}{\textit{\textbf{Large}}} &
  \multirow{2}{*}{\textit{\textbf{Dense}}} &
  \textit{\textbf{True}} &
  Logistic &
  0.444 &
  0.73 &
  0.99 \\ \cline{4-8} 
 &
   &
   &
  \textit{\textbf{False}} &
  Normal &
  0.37 &
  1.388 &
  - \\ \cline{3-8} 
 &
   &
  \multirow{2}{*}{\textit{\textbf{Wide}}} &
  \textit{\textbf{True}} &
  Logistic &
  0.57 &
  0.725 &
  0.689 \\ \cline{4-8} 
 &
   &
   &
  \textit{\textbf{False}} &
  Normal &
  0.518 &
  1.433 &
  - \\ \cline{2-8} 
 &
  \multirow{4}{*}{\textit{\textbf{Small}}} &
  \multirow{2}{*}{\textit{\textbf{Dense}}} &
  \textit{\textbf{True}} &
  Logistic &
  0.193 &
  0.581 &
  0.986 \\ \cline{4-8} 
 &
   &
   &
  \textit{\textbf{False}} &
  Normal &
  0.383 &
  1.292 &
  - \\ \cline{3-8} 
 &
   &
  \multirow{2}{*}{\textit{\textbf{Wide}}} &
  \textit{\textbf{True}} &
  Logistic &
  0.484 &
  0.653 &
  0.441 \\ \cline{4-8} 
 &
   &
   &
  \textit{\textbf{False}} &
  Normal &
  0.569 &
  1.397 &
  - \\ \hline
\end{tabular}
\end{table}

\section{SVM Model}

To build the Bayesian-based ML model, we used an SVM model, one of the ML models. The SVM model distinguishes the user's selection intention by utilizing posterior data calculated through the Bayesian model. The construction, training, and evaluation of the SVM model were done using Accord.Net, a C\# framework, which was chosen for its ease of integration with the C\#-based Unity environment. The SVM model employed is a Gaussian kernel-based SVM model, which infers binary classification results. The SVM model's hyperparameters were optimized through a simple grid search on the validation set. To prevent overfitting, 5-fold cross-validation was performed. In our training phase, we set the hyperparameters $\sigma$ = 30, Complexity = 30. Additionally, the dataset was randomly shuffled and sampled to maintain a 1:3 ratio for the class type based on selection, then split into training, validation, and test sets at a 6:2:2 ratio.

\section{User Study}

This section introduces detailed information about the first user study conducted for data collection and Study 2, which was designed to evaluate the gaze-based interaction technique proposed in this research. First, the demographic information of participants from both user studies is presented. Then, the statistical analysis results from Study 2 are provided in detail.

\subsection{Participants}

In Study 1, data collection was conducted with 20 participants. Their ages had a mean of 24.35 years with a standard deviation of 2.032, ranging from 21 to 30 years. The gender distribution included 11 males and 0 females. Study 2 involved 23 participants, with an average age of 25.652 years and a standard deviation of 6.383. The gender distribution in Study 2 included 8 males and 15 females.

\subsection{Statistical Analysis}

Study 2 employed a $2 \times 3$ within-subject factorial design, with target configuration and interaction technique as factors in the user study. The six measured dependent variables were SUS score, NASA-TLX overall score, NASA-TLX physical demand score, TTC, Error Rate, and user preference rank score. The analysis method is explained in the respective section, and the results of the analysis are presented in the Table~\ref{tab:stat:ttc} to~\ref{tab:stat:rank}.

\begin{table}[ht]
\caption{The results of RM-ANOVA and post-hoc tests for the TTC metric are presented. A two-way RM-ANOVA was conducted to analyze the interaction effect and main effects of Target Configuration and Interaction Technique. Additionally, a one-way RM-ANOVA was performed to assess the effects of Interaction Technique, with Bonferroni correction applied in the post-hoc tests. Statistically significant differences with p-values less than 0.05 are indicated by *.}
\label{tab:stat:ttc}
\resizebox{\textwidth}{!}{%
\begin{tabular}{c|clll|clll}
\hline
\multirow{2}{*}{\textbf{\begin{tabular}[c]{@{}c@{}}Independent\\ Variables\end{tabular}}} &
  \multicolumn{4}{c|}{\textbf{RM-ANOVA}} &
  \multicolumn{4}{c}{\textbf{post-hoc test}} \\ \cline{2-9} 
 &
  \multicolumn{1}{c|}{\textbf{Effects or Factors}} &
  \multicolumn{1}{c|}{\textit{\textbf{F-value}}} &
  \multicolumn{1}{c|}{\textit{\textbf{p-value}}} &
  \multicolumn{1}{c|}{\textbf{Effect Size}} &
  \multicolumn{1}{c|}{\textbf{Compairson}} &
  \multicolumn{1}{c|}{\textit{\textbf{t}}} &
  \multicolumn{1}{c|}{\textit{\textbf{p-value}}} &
  \multicolumn{1}{c}{\textit{\textbf{Cohen's d}}} \\ \hline
\multirow{3}{*}{\textbf{\begin{tabular}[c]{@{}c@{}}Target Configuration\\ $\times$\\ Interaction Technique\end{tabular}}} &
  \multicolumn{1}{c|}{\textbf{\begin{tabular}[c]{@{}c@{}}Interaction\\ Effects\end{tabular}}} &
  \multicolumn{1}{l|}{(1.457, 29.133) =   0.341} &
  \multicolumn{1}{l|}{0.646} &
  0.017 &
  \multicolumn{4}{c}{\multirow{3}{*}{-}} \\ \cline{2-5}
 &
  \multicolumn{1}{c|}{\textbf{\begin{tabular}[c]{@{}c@{}}Main Effects:\\ Target Configuration\end{tabular}}} &
  \multicolumn{1}{l|}{(1, 20) = 49.746} &
  \multicolumn{1}{l|}{\textless 0.001*} &
  0.713 &
  \multicolumn{4}{c}{} \\ \cline{2-5}
 &
  \multicolumn{1}{c|}{\textbf{\begin{tabular}[c]{@{}c@{}}Main Effects:\\ Interaction Technique\end{tabular}}} &
  \multicolumn{1}{l|}{(2, 40) = 0.651} &
  \multicolumn{1}{l|}{0.527} &
  0.032 &
  \multicolumn{4}{c}{} \\ \hline
\multirow{6}{*}{\textbf{Interaction Technique}} &
  \multicolumn{1}{c|}{\multirow{3}{*}{\textbf{Large $\times$ Dense}}} &
  \multicolumn{1}{l|}{\multirow{3}{*}{(2, 40) = 0.694}} &
  \multicolumn{1}{l|}{\multirow{3}{*}{0.505}} &
  \multirow{3}{*}{0.034} &
  \multicolumn{1}{c|}{-} &
  \multicolumn{1}{l|}{1.331} &
  \multicolumn{1}{l|}{0.595} &
  0.267 \\ \cline{6-9} 
 &
  \multicolumn{1}{c|}{} &
  \multicolumn{1}{l|}{} &
  \multicolumn{1}{l|}{} &
   &
  \multicolumn{1}{c|}{-} &
  \multicolumn{1}{l|}{0.092} &
  \multicolumn{1}{l|}{1} &
  0.027 \\ \cline{6-9} 
 &
  \multicolumn{1}{c|}{} &
  \multicolumn{1}{l|}{} &
  \multicolumn{1}{l|}{} &
   &
  \multicolumn{1}{c|}{-} &
  \multicolumn{1}{l|}{0.977} &
  \multicolumn{1}{l|}{1} &
  0.24 \\ \cline{2-9} 
 &
  \multicolumn{1}{c|}{\multirow{3}{*}{\textbf{Small $\times$ Wide}}} &
  \multicolumn{1}{l|}{\multirow{3}{*}{(2, 44) = 1.083}} &
  \multicolumn{1}{l|}{\multirow{3}{*}{0.347}} &
  \multirow{3}{*}{0.047} &
  \multicolumn{1}{c|}{-} &
  \multicolumn{1}{l|}{0.364} &
  \multicolumn{1}{l|}{1} &
  0.052 \\ \cline{6-9} 
 &
  \multicolumn{1}{c|}{} &
  \multicolumn{1}{l|}{} &
  \multicolumn{1}{l|}{} &
   &
  \multicolumn{1}{c|}{-} &
  \multicolumn{1}{l|}{1.261} &
  \multicolumn{1}{l|}{0.662} &
  0.207 \\ \cline{6-9} 
 &
  \multicolumn{1}{c|}{} &
  \multicolumn{1}{l|}{} &
  \multicolumn{1}{l|}{} &
   &
  \multicolumn{1}{c|}{-} &
  \multicolumn{1}{l|}{1.201} &
  \multicolumn{1}{l|}{0.727} &
  0.155 \\ \hline
\end{tabular}%
}
\end{table}

\begin{table}[ht]
\caption{The results of RM-ANOVA and post-hoc tests for the Error Rate metric are presented. A two-way RM-ANOVA was conducted to analyze the interaction effect and main effects of Target Configuration and Interaction Technique. Additionally, a one-way RM-ANOVA was performed to assess the effects of Interaction Technique, with Bonferroni correction applied in the post-hoc tests. Statistically significant differences with p-values less than 0.05 are indicated by *.}
\label{tab:stat:err}
\resizebox{\textwidth}{!}{%
\begin{tabular}{c|clll|clll}
\hline
\multirow{2}{*}{\textbf{\begin{tabular}[c]{@{}c@{}}Independent\\ Variables\end{tabular}}} &
  \multicolumn{4}{c|}{\textbf{RM-ANOVA}} &
  \multicolumn{4}{c}{\textbf{post-hoc test}} \\ \cline{2-9} 
 &
  \multicolumn{1}{c|}{\textbf{Effects or Factors}} &
  \multicolumn{1}{c|}{\textit{\textbf{F-value}}} &
  \multicolumn{1}{c|}{\textit{\textbf{p-value}}} &
  \multicolumn{1}{c|}{\textbf{Effect Size}} &
  \multicolumn{1}{c|}{\textbf{Compairson}} &
  \multicolumn{1}{c|}{\textit{\textbf{t}}} &
  \multicolumn{1}{c|}{\textit{\textbf{p-value}}} &
  \multicolumn{1}{c}{\textit{\textbf{Cohen's d}}} \\ \hline
\multirow{3}{*}{\textbf{\begin{tabular}[c]{@{}c@{}}Target Configuration\\ $\times$\\ Interaction Technique\end{tabular}}} &
  \multicolumn{1}{c|}{\textbf{\begin{tabular}[c]{@{}c@{}}Interaction\\ Effects\end{tabular}}} &
  \multicolumn{1}{l|}{(2, 40) = 8.47} &
  \multicolumn{1}{l|}{\textless 0.001*} &
  0.294 &
  \multicolumn{4}{c}{\multirow{3}{*}{-}} \\ \cline{2-5}
 &
  \multicolumn{1}{c|}{\textbf{\begin{tabular}[c]{@{}c@{}}Main Effects:\\ Target Configuration\end{tabular}}} &
  \multicolumn{1}{l|}{(1, 20) = 110.784} &
  \multicolumn{1}{l|}{\textless 0.001*} &
  0.847 &
  \multicolumn{4}{c}{} \\ \cline{2-5}
 &
  \multicolumn{1}{c|}{\textbf{\begin{tabular}[c]{@{}c@{}}Main Effects:\\ Interaction Technique\end{tabular}}} &
  \multicolumn{1}{l|}{(2, 40) = 32.151} &
  \multicolumn{1}{l|}{\textless 0.001*} &
  0.616 &
  \multicolumn{4}{c}{} \\ \hline
\multirow{6}{*}{\textbf{Interaction Technique}} &
  \multicolumn{1}{c|}{\multirow{3}{*}{\textbf{Large $\times$ Dense}}} &
  \multicolumn{1}{l|}{\multirow{3}{*}{(1.237, 24.737) = 21.28}} &
  \multicolumn{1}{l|}{\multirow{3}{*}{\textless 0.001*}} &
  \multirow{3}{*}{0.516} &
  \multicolumn{1}{c|}{Gaze \textless Cont} &
  \multicolumn{1}{l|}{5.933} &
  \multicolumn{1}{l|}{\textless 0.001*} &
  1.835 \\ \cline{6-9} 
 &
  \multicolumn{1}{c|}{} &
  \multicolumn{1}{l|}{} &
  \multicolumn{1}{l|}{} &
   &
  \multicolumn{1}{c|}{-} &
  \multicolumn{1}{l|}{0.993} &
  \multicolumn{1}{l|}{0.998} &
  0.175 \\ \cline{6-9} 
 &
  \multicolumn{1}{c|}{} &
  \multicolumn{1}{l|}{} &
  \multicolumn{1}{l|}{} &
   &
  \multicolumn{1}{c|}{Cont \textgreater Dwell} &
  \multicolumn{1}{l|}{4.12} &
  \multicolumn{1}{l|}{0.002*} &
  1.66 \\ \cline{2-9} 
 &
  \multicolumn{1}{c|}{\multirow{3}{*}{\textbf{Small $\times$ Wide}}} &
  \multicolumn{1}{l|}{\multirow{3}{*}{(2, 44) = 20.372}} &
  \multicolumn{1}{l|}{\multirow{3}{*}{\textless 0.001*}} &
  \multirow{3}{*}{0.481} &
  \multicolumn{1}{c|}{Gaze \textless Cont} &
  \multicolumn{1}{l|}{6.372} &
  \multicolumn{1}{l|}{\textless 0.001*} &
  1.727 \\ \cline{6-9} 
 &
  \multicolumn{1}{c|}{} &
  \multicolumn{1}{l|}{} &
  \multicolumn{1}{l|}{} &
   &
  \multicolumn{1}{c|}{Gaze \textless Dwell} &
  \multicolumn{1}{l|}{3.798} &
  \multicolumn{1}{l|}{0.003*} &
  0.929 \\ \cline{6-9} 
 &
  \multicolumn{1}{c|}{} &
  \multicolumn{1}{l|}{} &
  \multicolumn{1}{l|}{} &
   &
  \multicolumn{1}{c|}{Cont \textgreater Dwell} &
  \multicolumn{1}{l|}{2.709} &
  \multicolumn{1}{l|}{0.038*} &
  0.798 \\ \hline
\end{tabular}%
}
\end{table}

\newpage

\begin{table}[ht]
\caption{The results of RM-ANOVA and post-hoc tests for the NASA-TLX overall score metric are presented. A two-way RM-ANOVA was conducted to analyze the interaction effect and main effects of Target Configuration and Interaction Technique. Additionally, a one-way RM-ANOVA was performed to assess the effects of Interaction Technique, with Bonferroni correction applied in the post-hoc tests. Statistically significant differences with p-values less than 0.05 are indicated by *.}
\label{tab:stat:nasaoverall}
\resizebox{\textwidth}{!}{%
\begin{tabular}{c|clll|clll}
\hline
\multirow{2}{*}{\textbf{\begin{tabular}[c]{@{}c@{}}Independent\\ Variables\end{tabular}}} &
  \multicolumn{4}{c|}{\textbf{RM-ANOVA}} &
  \multicolumn{4}{c}{\textbf{post-hoc test}} \\ \cline{2-9} 
 &
  \multicolumn{1}{c|}{\textbf{Effects or Factors}} &
  \multicolumn{1}{c|}{\textit{\textbf{F-value}}} &
  \multicolumn{1}{c|}{\textit{\textbf{p-value}}} &
  \multicolumn{1}{c|}{\textbf{Effect Size}} &
  \multicolumn{1}{c|}{\textbf{Compairson}} &
  \multicolumn{1}{c|}{\textit{\textbf{t}}} &
  \multicolumn{1}{c|}{\textit{\textbf{p-value}}} &
  \multicolumn{1}{c}{\textit{\textbf{Cohen's d}}} \\ \hline
\multirow{3}{*}{\textbf{\begin{tabular}[c]{@{}c@{}}Target Configuration\\ $\times$\\ Interaction Technique\end{tabular}}} &
  \multicolumn{1}{c|}{\textbf{\begin{tabular}[c]{@{}c@{}}Interaction\\ Effects\end{tabular}}} &
  \multicolumn{1}{l|}{(2, 44) = 5.423} &
  \multicolumn{1}{l|}{0.008*} &
  0.198 &
  \multicolumn{4}{c}{\multirow{3}{*}{-}} \\ \cline{2-5}
 &
  \multicolumn{1}{c|}{\textbf{\begin{tabular}[c]{@{}c@{}}Main Effects:\\ Target Configuration\end{tabular}}} &
  \multicolumn{1}{l|}{(1, 22) = 2.502} &
  \multicolumn{1}{l|}{0.128} &
  0.102 &
  \multicolumn{4}{c}{} \\ \cline{2-5}
 &
  \multicolumn{1}{c|}{\textbf{\begin{tabular}[c]{@{}c@{}}Main Effects:\\ Interaction Technique\end{tabular}}} &
  \multicolumn{1}{l|}{(1.472, 32.386) = 5.313} &
  \multicolumn{1}{l|}{0.017*} &
  0.195 &
  \multicolumn{4}{c}{} \\ \hline
\multirow{6}{*}{\textbf{Interaction Technique}} &
  \multicolumn{1}{c|}{\multirow{3}{*}{\textbf{Large $\times$ Dense}}} &
  \multicolumn{1}{l|}{\multirow{3}{*}{(2, 44) = 0.024}} &
  \multicolumn{1}{l|}{\multirow{3}{*}{0.976}} &
  \multirow{3}{*}{0.001} &
  \multicolumn{1}{c|}{-} &
  \multicolumn{1}{l|}{0.218} &
  \multicolumn{1}{l|}{1} &
  0.026 \\ \cline{6-9} 
 &
  \multicolumn{1}{c|}{} &
  \multicolumn{1}{l|}{} &
  \multicolumn{1}{l|}{} &
   &
  \multicolumn{1}{c|}{-} &
  \multicolumn{1}{l|}{0.196} &
  \multicolumn{1}{l|}{1} &
  0.005 \\ \cline{6-9} 
 &
  \multicolumn{1}{c|}{} &
  \multicolumn{1}{l|}{} &
  \multicolumn{1}{l|}{} &
   &
  \multicolumn{1}{c|}{-} &
  \multicolumn{1}{l|}{0.035} &
  \multicolumn{1}{l|}{1} &
  0.021 \\ \cline{2-9} 
 &
  \multicolumn{1}{c|}{\multirow{3}{*}{\textbf{Small $\times$ Wide}}} &
  \multicolumn{1}{l|}{\multirow{3}{*}{(1.517, 33.367) = 10.139}} &
  \multicolumn{1}{l|}{\multirow{3}{*}{\textless 0.001*}} &
  \multirow{3}{*}{0.315} &
  \multicolumn{1}{c|}{Gaze \textless Cont} &
  \multicolumn{1}{l|}{2.871} &
  \multicolumn{1}{l|}{0.027*} &
  0.555 \\ \cline{6-9} 
 &
  \multicolumn{1}{c|}{} &
  \multicolumn{1}{l|}{} &
  \multicolumn{1}{l|}{} &
   &
  \multicolumn{1}{c|}{Gaze \textless Dwell} &
  \multicolumn{1}{l|}{5.750} &
  \multicolumn{1}{l|}{\textless 0.001*} &
  0.660 \\ \cline{6-9} 
 &
  \multicolumn{1}{c|}{} &
  \multicolumn{1}{l|}{} &
  \multicolumn{1}{l|}{} &
   &
  \multicolumn{1}{c|}{-} &
  \multicolumn{1}{l|}{0.679} &
  \multicolumn{1}{l|}{1} &
  0.105 \\ \hline
\end{tabular}%
}
\end{table}

\begin{table}[ht]
\caption{The results of RM-ANOVA and post-hoc tests for the NASA-TLX physical demand score metric are presented. A two-way RM-ANOVA was conducted to analyze the interaction effect and main effects of Target Configuration and Interaction Technique. Additionally, a one-way RM-ANOVA was performed to assess the effects of Interaction Technique, with Bonferroni correction applied in the post-hoc tests. Statistically significant differences with p-values less than 0.05 are indicated by *.}
\label{tab:stat:nasaphy}
\resizebox{\textwidth}{!}{%
\begin{tabular}{c|clll|clll}
\hline
\multirow{2}{*}{\textbf{\begin{tabular}[c]{@{}c@{}}Independent\\ Variables\end{tabular}}} &
  \multicolumn{4}{c|}{\textbf{RM-ANOVA}} &
  \multicolumn{4}{c}{\textbf{post-hoc test}} \\ \cline{2-9} 
 &
  \multicolumn{1}{c|}{\textbf{Effects or Factors}} &
  \multicolumn{1}{c|}{\textit{\textbf{F-value}}} &
  \multicolumn{1}{c|}{\textit{\textbf{p-value}}} &
  \multicolumn{1}{c|}{\textbf{Effect Size}} &
  \multicolumn{1}{c|}{\textbf{Compairson}} &
  \multicolumn{1}{c|}{\textit{\textbf{t}}} &
  \multicolumn{1}{c|}{\textit{\textbf{p-value}}} &
  \multicolumn{1}{c}{\textit{\textbf{Cohen's d}}} \\ \hline
\multirow{3}{*}{\textbf{\begin{tabular}[c]{@{}c@{}}Target Configuration\\ $\times$\\ Interaction Technique\end{tabular}}} &
  \multicolumn{1}{c|}{\textbf{\begin{tabular}[c]{@{}c@{}}Interaction\\ Effects\end{tabular}}} &
  \multicolumn{1}{l|}{(2, 40) = 3.672} &
  \multicolumn{1}{l|}{0.034*} &
  0.155 &
  \multicolumn{4}{c}{\multirow{3}{*}{-}} \\ \cline{2-5}
 &
  \multicolumn{1}{c|}{\textbf{\begin{tabular}[c]{@{}c@{}}Main Effects:\\ Target Configuration\end{tabular}}} &
  \multicolumn{1}{l|}{(1, 20) = 14.085} &
  \multicolumn{1}{l|}{0.001*} &
  0.413 &
  \multicolumn{4}{c}{} \\ \cline{2-5}
 &
  \multicolumn{1}{c|}{\textbf{\begin{tabular}[c]{@{}c@{}}Main Effects:\\ Interaction Technique\end{tabular}}} &
  \multicolumn{1}{l|}{(1.165, 23.31) = 8.854} &
  \multicolumn{1}{l|}{0.005*} &
  0.307 &
  \multicolumn{4}{c}{} \\ \hline
\multirow{6}{*}{\textbf{Interaction Technique}} &
  \multicolumn{1}{c|}{\multirow{3}{*}{\textbf{Large $\times$ Dense}}} &
  \multicolumn{1}{l|}{\multirow{3}{*}{(1.173, 23.456) = 6.056}} &
  \multicolumn{1}{l|}{\multirow{3}{*}{0.018*}} &
  \multirow{3}{*}{0.232} &
  \multicolumn{1}{c|}{-} &
  \multicolumn{1}{l|}{2.517} &
  \multicolumn{1}{l|}{0.061} &
  0.72 \\ \cline{6-9} 
 &
  \multicolumn{1}{c|}{} &
  \multicolumn{1}{l|}{} &
  \multicolumn{1}{l|}{} &
   &
  \multicolumn{1}{c|}{-} &
  \multicolumn{1}{l|}{0.623} &
  \multicolumn{1}{l|}{1} &
  0.06 \\ \cline{6-9} 
 &
  \multicolumn{1}{c|}{} &
  \multicolumn{1}{l|}{} &
  \multicolumn{1}{l|}{} &
   &
  \multicolumn{1}{c|}{-} &
  \multicolumn{1}{l|}{2.549} &
  \multicolumn{1}{l|}{0.057} &
  0.66 \\ \cline{2-9} 
 &
  \multicolumn{1}{c|}{\multirow{3}{*}{\textbf{Small $\times$ Wide}}} &
  \multicolumn{1}{l|}{\multirow{3}{*}{(1.436, 31.581) = 11.322}} &
  \multicolumn{1}{l|}{\multirow{3}{*}{\textless 0.001*}} &
  \multirow{3}{*}{0.34} &
  \multicolumn{1}{c|}{Gaze \textless Cont} &
  \multicolumn{1}{l|}{4.033} &
  \multicolumn{1}{l|}{0.002*} &
  1.043 \\ \cline{6-9} 
 &
  \multicolumn{1}{c|}{} &
  \multicolumn{1}{l|}{} &
  \multicolumn{1}{l|}{} &
   &
  \multicolumn{1}{c|}{Gaze \textless Dwell} &
  \multicolumn{1}{l|}{4.72} &
  \multicolumn{1}{l|}{\textless 0.001*} &
  0.64 \\ \cline{6-9} 
 &
  \multicolumn{1}{c|}{} &
  \multicolumn{1}{l|}{} &
  \multicolumn{1}{l|}{} &
   &
  \multicolumn{1}{c|}{-} &
  \multicolumn{1}{l|}{1.627} &
  \multicolumn{1}{l|}{0.354} &
  0.403 \\ \hline
\end{tabular}%
}
\end{table}

\newpage

\begin{table}[ht]
\caption{The results of RM-ANOVA and post-hoc tests for the SUS score metric are presented. A two-way RM-ANOVA was conducted to analyze the interaction effect and main effects of Target Configuration and Interaction Technique. Additionally, a one-way RM-ANOVA was performed to assess the effects of Interaction Technique, with Bonferroni correction applied in the post-hoc tests. Statistically significant differences with p-values less than 0.05 are indicated by *.}
\label{tab:stat:sus}
\resizebox{\textwidth}{!}{%
\begin{tabular}{c|clll|clll}
\hline
\multirow{2}{*}{\textbf{\begin{tabular}[c]{@{}c@{}}Independent\\ Variables\end{tabular}}} &
  \multicolumn{4}{c|}{\textbf{RM-ANOVA}} &
  \multicolumn{4}{c}{\textbf{post-hoc test}} \\ \cline{2-9} 
 &
  \multicolumn{1}{c|}{\textbf{Effects or Factors}} &
  \multicolumn{1}{c|}{\textit{\textbf{F-value}}} &
  \multicolumn{1}{c|}{\textit{\textbf{p-value}}} &
  \multicolumn{1}{c|}{\textbf{Effect Size}} &
  \multicolumn{1}{c|}{\textbf{Compairson}} &
  \multicolumn{1}{c|}{\textit{\textbf{t}}} &
  \multicolumn{1}{c|}{\textit{\textbf{p-value}}} &
  \multicolumn{1}{c}{\textit{\textbf{Cohen's d}}} \\ \hline
\multirow{3}{*}{\textbf{\begin{tabular}[c]{@{}c@{}}Target Configuration\\ $\times$\\ Interaction Technique\end{tabular}}} &
  \multicolumn{1}{c|}{\textbf{\begin{tabular}[c]{@{}c@{}}Interaction\\ Effects\end{tabular}}} &
  \multicolumn{1}{l|}{(2, 40) = 0.517} &
  \multicolumn{1}{l|}{0.6} &
  0.025 &
  \multicolumn{4}{c}{\multirow{3}{*}{-}} \\ \cline{2-5}
 &
  \multicolumn{1}{c|}{\textbf{\begin{tabular}[c]{@{}c@{}}Main Effects:\\ Target Configuration\end{tabular}}} &
  \multicolumn{1}{l|}{(1, 20) = 1.485} &
  \multicolumn{1}{l|}{0.237} &
  0.069 &
  \multicolumn{4}{c}{} \\ \cline{2-5}
 &
  \multicolumn{1}{c|}{\textbf{\begin{tabular}[c]{@{}c@{}}Main Effects:\\ Interaction Technique\end{tabular}}} &
  \multicolumn{1}{l|}{(2, 40) = 1.24} &
  \multicolumn{1}{l|}{0.3} &
  0.058 &
  \multicolumn{4}{c}{} \\ \hline
\multirow{6}{*}{\textbf{Interaction Technique}} &
  \multicolumn{1}{c|}{\multirow{3}{*}{\textbf{Large $\times$ Dense}}} &
  \multicolumn{1}{l|}{\multirow{3}{*}{(2, 40) = 1.570}} &
  \multicolumn{1}{l|}{\multirow{3}{*}{0.221}} &
  \multirow{3}{*}{0.073} &
  \multicolumn{1}{c|}{-} &
  \multicolumn{1}{l|}{1.56} &
  \multicolumn{1}{l|}{0.403} &
  0.291 \\ \cline{6-9} 
 &
  \multicolumn{1}{c|}{} &
  \multicolumn{1}{l|}{} &
  \multicolumn{1}{l|}{} &
   &
  \multicolumn{1}{c|}{-} &
  \multicolumn{1}{l|}{0.228} &
  \multicolumn{1}{l|}{1} &
  0.046 \\ \cline{6-9} 
 &
  \multicolumn{1}{c|}{} &
  \multicolumn{1}{l|}{} &
  \multicolumn{1}{l|}{} &
   &
  \multicolumn{1}{c|}{-} &
  \multicolumn{1}{l|}{1.83} &
  \multicolumn{1}{l|}{0.247} &
  0.245 \\ \cline{2-9} 
 &
  \multicolumn{1}{c|}{\multirow{3}{*}{\textbf{Small $\times$ Wide}}} &
  \multicolumn{1}{l|}{\multirow{3}{*}{(2, 44) = 0.146}} &
  \multicolumn{1}{l|}{\multirow{3}{*}{0.865}} &
  \multirow{3}{*}{0.007} &
  \multicolumn{1}{c|}{-} &
  \multicolumn{1}{l|}{0.488} &
  \multicolumn{1}{l|}{1} &
  0.09 \\ \cline{6-9} 
 &
  \multicolumn{1}{c|}{} &
  \multicolumn{1}{l|}{} &
  \multicolumn{1}{l|}{} &
   &
  \multicolumn{1}{c|}{-} &
  \multicolumn{1}{l|}{0.419} &
  \multicolumn{1}{l|}{1} &
  0.068 \\ \cline{6-9} 
 &
  \multicolumn{1}{c|}{} &
  \multicolumn{1}{l|}{} &
  \multicolumn{1}{l|}{} &
   &
  \multicolumn{1}{c|}{-} &
  \multicolumn{1}{l|}{0.129} &
  \multicolumn{1}{l|}{1} &
  0.023 \\ \hline
\end{tabular}%
}
\end{table}

\begin{table}[ht]
\caption{The results of RM-ANOVA and post-hoc tests for the User Preference rank score metric are presented. A one-way RM-ANOVA was performed to assess the effects of Interaction Technique, with Bonferroni correction applied in the post-hoc tests. Statistically significant differences with p-values less than 0.05 are indicated by *.}
\label{tab:stat:rank}
\resizebox{\textwidth}{!}{%
\begin{tabular}{c|clll|clll}
\hline
\multirow{2}{*}{\textbf{\begin{tabular}[c]{@{}c@{}}Independent\\ Variables\end{tabular}}} &
  \multicolumn{4}{c|}{\textbf{RM-ANOVA}} &
  \multicolumn{4}{c}{\textbf{post-hoc test}} \\ \cline{2-9} 
 &
  \multicolumn{1}{c|}{\textbf{Effects or Factors}} &
  \multicolumn{1}{c|}{\textit{\textbf{F-value}}} &
  \multicolumn{1}{c|}{\textit{\textbf{p-value}}} &
  \multicolumn{1}{c|}{\textbf{Effect Size}} &
  \multicolumn{1}{c|}{\textbf{Compairson}} &
  \multicolumn{1}{c|}{\textit{\textbf{t}}} &
  \multicolumn{1}{c|}{\textit{\textbf{p-value}}} &
  \multicolumn{1}{c}{\textit{\textbf{Cohen's d}}} \\ \hline
\multirow{6}{*}{\textbf{Interaction Technique}} &
  \multicolumn{1}{c|}{\multirow{3}{*}{\textbf{Large $\times$ Dense}}} &
  \multicolumn{1}{l|}{\multirow{3}{*}{2 = 3.391}} &
  \multicolumn{1}{l|}{\multirow{3}{*}{0.183}} &
  \multirow{3}{*}{0.074} &
  \multicolumn{1}{c|}{-} &
  \multicolumn{1}{l|}{1.348} &
  \multicolumn{1}{l|}{0.553} &
   \\ \cline{6-9} 
 &
  \multicolumn{1}{c|}{} &
  \multicolumn{1}{l|}{} &
  \multicolumn{1}{l|}{} &
   &
  \multicolumn{1}{c|}{-} &
  \multicolumn{1}{l|}{0.449} &
  \multicolumn{1}{l|}{1} &
   \\ \cline{6-9} 
 &
  \multicolumn{1}{c|}{} &
  \multicolumn{1}{l|}{} &
  \multicolumn{1}{l|}{} &
   &
  \multicolumn{1}{c|}{-} &
  \multicolumn{1}{l|}{1.798} &
  \multicolumn{1}{l|}{0.237} &
   \\ \cline{2-9} 
 &
  \multicolumn{1}{c|}{\multirow{3}{*}{\textbf{Small $\times$ Wide}}} &
  \multicolumn{1}{l|}{\multirow{3}{*}{2 = 7.913}} &
  \multicolumn{1}{l|}{\multirow{3}{*}{0.019*}} &
  \multirow{3}{*}{0.172} &
  \multicolumn{1}{c|}{-} &
  \multicolumn{1}{l|}{0.158} &
  \multicolumn{1}{l|}{1} &
   \\ \cline{6-9} 
 &
  \multicolumn{1}{c|}{} &
  \multicolumn{1}{l|}{} &
  \multicolumn{1}{l|}{} &
   &
  \multicolumn{1}{l|}{Gaze \textless Dwell} &
  \multicolumn{1}{l|}{2.694} &
  \multicolumn{1}{l|}{0.03*} &
   \\ \cline{6-9} 
 &
  \multicolumn{1}{c|}{} &
  \multicolumn{1}{l|}{} &
  \multicolumn{1}{l|}{} &
   &
  \multicolumn{1}{l|}{Cont \textless Dwell} &
  \multicolumn{1}{l|}{2.536} &
  \multicolumn{1}{l|}{0.045*} &
   \\ \hline
\end{tabular}%
}
\end{table}

\bibliographystyle{ACM-Reference-Format}
\bibliography{SupplementaryMaterial}